\documentclass[11pt]{article}

\usepackage[a4paper,margin=1in]{geometry}
\usepackage[T1]{fontenc}
\usepackage{times}
\usepackage{graphicx}
\usepackage{authblk}

\usepackage{moreverb,url}
\usepackage{amsthm}
\usepackage{amsmath}
\usepackage{amssymb}
\usepackage{enumitem}

\newtheorem{definition}{Definition}

\usepackage[colorlinks,bookmarksopen,bookmarksnumbered,citecolor=red,urlcolor=red]{hyperref}

\begin{document}

\title{DeceptionAnalyser: A Web-Based AI Tool for Performing Structured Deception Analysis with Argumentation Schemes and LLMs}

\author[1]{Stefan Sarkadi\thanks{Corresponding author: University of Lincoln, Brayford Way, Lincoln, LN6 7TS, UK.\ \texttt{ssarkadi@lincoln.ac.uk}}}
\author[2]{Xabier Garmendia}
\author[3]{Jack Mumford}
\author[3]{Trevor Bench-Capon}
\affil[1]{University of Lincoln, UK}
\affil[2]{University of the Basque Country, Spain}
\affil[3]{University of Liverpool, UK}
\date{}

\maketitle

\begin{abstract}
Deception plays a central role in Intelligence operations, yet it remains difficult to analyse systematically without expert knowledge of reasoning patterns and cognitive manipulation. In computational argumentation, for instance, no scheme-level ground-truth corpora currently exist to support statistical validation. In this paper, we address this by introducing a set of ten argument schemes designed to model distinct forms of deception, each accompanied by structured premises and critical questions. In doing so, we introduce the first dedicated library of argumentation schemes specifically designed for deception analysis, providing a structured foundation for systematically modelling and analysing deception in narrative text. We then present \textit{DeceptionAnalyser}, a browser-based tool that implements these schemes through a two-stage methodology combining LLM-based premise extraction with critical-question-driven evaluation. Our aim is to provide a conceptual and methodological foundation for analysing deceptive reasoning in narrative text. This is precisely what we address in this paper by demonstrating how structured argumentation theory and AI-assisted analysis can support transparent, explainable assessments of potential deception. Because the schemes are designed to flag claims for scrutiny rather than to output a deception verdict, we do not benchmark classification accuracy; instead, we assess the \emph{reliability} of the methodology by measuring the consistency of the tool's premise and conclusion assessments across ten contemporary large language models and repeated runs. We find that scheme detection is highly stable for clear-cut deception and degrades gracefully, in interpretable ways, on more ambiguous intelligence-style narratives.
\medskip
\noindent\textbf{Keywords:} Deception Analysis, Argumentation, LLM, Intelligence Analysis, Human-AI Teaming, Hybrid Intelligence
\end{abstract}

\section{Introduction}

Deception plays a central role in intelligence analysis and in many forms of strategic communication, yet systematic methods for identifying and evaluating deceptive reasoning remain underdeveloped within computational argumentation. While prior work has explored how agents may deceive using dishonest communication \cite{sakama2010many,sakama2010logical,Caminada-2009-LanBDCofD}, strategic dialogue \cite{sarkadi2019DecStoryAAAI,sakama2012dishonest}, belief--intention models \cite{sarkadi2019modelling}, and manipulative reasoning patterns \cite{sakama2011dishonest}, there is currently no structured library of argument schemes tailored specifically to modelling deception in narrative text. 

Our paper has two major research aims that it addresses with two respective main contributions. (I) The \textbf{first} is to introduce a comprehensive set of deception-oriented argument schemes grounded in argumentation theory, cognitive models of deception, and structured analytic techniques. By doing so, we provide the \textbf{first dedicated library of argumentation schemes specifically designed for deception analysis}, filling a gap in both computational argumentation and deception focused intelligence methodology, where no structured scheme-based catalogue previously existed. These schemes therefore establish the first formal foundation for systematically modelling deceptive reasoning across narrative contexts. (II) The \textbf{second} is to present the \textit{DeceptionAnalyser} tool, a practical methodological framework that applies these schemes using a two-stage pipeline. The pipeline combines LLM-based premise extraction \cite{haddadan2019disputool} with critical-question-driven evaluation \cite{walton2008argumentation}. Rather than benchmarking deception-classification accuracy against a ground-truth corpus --- which does not exist at the scheme level, and which the method is not designed to produce --- we pursue two complementary forms of validation. First, we provide a conceptual and methodological foundation for analysing deception, illustrating how the proposed schemes and tool support transparent, explainable, and human-guided reasoning about deception in narrative texts. Second, we report a reliability study that measures how consistently the pipeline instantiates and assesses the schemes across ten contemporary LLMs and across repeated runs, thereby addressing the stochasticity and model-dependence of LLM-based analysis.

Our proposed approach is in line with the desiderata of hybrid intelligence and Human-AI teaming for deception analysis described in \cite{sarkadi2024deception}. This approach has two key advantages: it is \textbf{explainable} because it shows the reasoning process and evidence behind its conclusions, and it is \textbf{robust} because it systematically compares multiple interpretations rather than simply detecting suspicious language patterns.

The remainder of this paper is organised as follows. In Section 2, we introduce the theoretical background underpinning our approach, including relevant work on argumentation schemes, deceptive reasoning, and structured analytic techniques. Section 3 presents our library of ten deception-oriented argument schemes, detailing their premises, conclusions, and associated critical questions. Section 4 describes the \textit{DeceptionAnalyser} tool and its two-stage analytical methodology, showing how the schemes are implemented through LLM-based premise extraction and critical-question-driven evaluation. Section 5 provides an illustrative analysis demonstrating how the tool can be applied to narrative text, together with a reliability study of the pipeline's consistency across multiple LLMs and repeated runs. Section 6 discusses the limitations of the tool. Section 7 concludes the paper. The full prompt templates used by the tool are provided in the Appendix.

\section{Background}

In this section we introduce the background concepts alongside related works from the literature. We focus on research on argumentation schemes, deceptive reasoning in artificial agents and communication, and structured analytic techniques used in intelligence analysis. We also review literature on human--AI teaming to contextualise how computational tools can support analysts in reasoning about deception.

\subsection{Argumentation}

Argumentation schemes constitute formalised abstractions of stereotypical reasoning patterns that capture the inferential structure of everyday arguments \cite{walton2008argumentation}. Each scheme identifies a prototypical relation between premises and a conclusion, together with a corresponding set of \textit{critical questions} that guide the evaluation of an argument's validity and possible weaknesses. Common examples include argument from expert opinion, argument from cause to effect, and argument from consequences \cite{macagno2017argumentation}. Within computational argumentation, schemes enable the mapping of natural language discourse onto formal representations, allowing automated systems to analyse and assess reasoning in a normatively interpretable manner, including assessing trust \cite{Parsons-2012-UAtoRWandAT,Parsons-2011-ABRinAwithVDofT}. Argument mining, which uses NLP techniques, has been also used for analysing dishonesty \cite{haddadan2019disputool}.

\subsection{Structured Analytic Techniques}

The psychology of intelligence analysis investigates the cognitive biases and decision-making processes that affect reasoning under uncertainty \cite{Heuer1999psychology,tversky1974judgment}. Intelligence analysts operate in contexts where information is incomplete, ambiguous, and sometimes deliberately misleading. These contexts make the analysts' jobs vulnerable to systematic cognitive biases such as confirmation bias, anchoring, and availability heuristics, which may negatively impact their intelligence reports. Examples include focusing on a premature hypothesis or overconfidence in flawed conclusions. Deception exploits precisely these cognitive tendencies by offering coherent but misleading information that aligns with an analyst's expectations, or by concealing salient information to maximise the state of unknown unknowns.

Structured Analytic Techniques (SATs) are systematic methods developed to mitigate these effects \cite{pherson2019structured,dhami2019analysis}. Techniques such as Analysis of Competing Hypotheses (ACH), Key Assumptions Check, and Red Team Analysis are designed to counteract cognitive bias, facilitate the explicit mapping of evidence to hypotheses, and support reproducible decision-making processes. ACH, in particular, is well-suited to computational adaptation because it formalises reasoning as a matrix of evidential consistency and inconsistency across competing hypotheses, encouraging analysts to adopt a probabilistic and falsification-oriented mindset. Another complementary method to ACH is subjective logic, that includes the assessment of trust in multiple dimensions, including trust in different sources \cite{Josang-1997-ARwithSL,pope2005analysis}

For instance, ACH is done in a continuous cycle of seven iterative steps. The seven steps are:

\begin{enumerate}[leftmargin=*]
    \item \textbf{Hypothesis}: Analysts brainstorm in order to exhaust all possible hypotheses for a given case.
    \item \textbf{Evidence}: Analysts bring evidence (arguments and data) for and against every hypothesis they came up with in the previous step.
    \item \textbf{Diagnostics}: In this step, the analysts attempt to disprove as many hypotheses as possible by using evidence that falsifies the hypotheses. One important factor is the diagnosticity of evidence. The higher the diagnosticity of a piece of evidence, the more likely it is for that piece of evidence to disprove a hypothesis. This is similar to the falsification principle in the scientific method \cite{Popper2005logic}.
    \item \textbf{Refinement}: After the first three steps, the analysts proceed to refine their finding by identifying additional evidence to refute as many hypotheses as possible.
    \item \textbf{Inconsistency}: The analysts aim to find and solve any inconsistencies between the hypotheses, drawing tentative conclusions about the likelihood of hypotheses.
    \item \textbf{Sensitivity}: In this step, the analysts try to estimate what impact their assumptions and evidence would have on their conclusions if they were proved to be false.
    \item \textbf{Conclusion and evaluation}: The analysts provide the conclusion of their analysis along with an evaluation of each step they have taken in the analysis. The analysts must also include in the evaluation the alternatives they had to reject during this process.
\end{enumerate}

\subsection{Human--AI Teaming in Intelligence Analysis}

In the context of Intelligence Analysis, the paradigm of human--AI teaming focuses on the collaboration between human analysts and intelligent systems \cite{endsley2024humans}. In the context of deception detection, effective teaming harnesses the complementary strengths of human intuition, contextual reasoning, and ethical judgement alongside AI's ability to process and cross-check large volumes of information. The success of such collaboration depends critically on explainability, transparency, and bidirectional interpretability: the AI must present its reasoning in forms accessible to human analysts, and humans must be able to critique, adapt, and refine machine outputs \cite{toniolo2023human}.

The integration of argumentation schemes and ACH provides a natural interface for such collaboration \cite{jones2018critical}. Argument structures make the reasoning process explicit and interpretable, while ACH formalism supports collaborative hypothesis testing through structured evidential analysis. Together, they facilitate interactive sensemaking, allowing analysts to explore the AI's argument maps, challenge assumptions, and iteratively refine hypotheses \cite{van2008can}.

An important challenge of human--machine interaction arises in environments where both humans and AI systems are potential agents of deception  \cite{sarkadi2024self}. Hybrid human-AI organisations require transparent architectures and protocols capable of modelling not only human beliefs but also machine intentions and communicative acts. The research on formal mechanisms for trust calibration and for detecting deceptive behaviour in mixed human--AI environments are crucial for enhancing mutual interpretability and accountability \cite{patel2025give}.

\subsection{AI-driven Deception Analysis}

Regarding deception detection, argumentation schemes offer an analytical mechanism for identifying manipulative or fallacious reasoning \cite{PanissonSarkadi-2018-LBandDinAOPL}. Deceptive narratives often imitate valid argumentative structures but introduce subtle distortions --- such as suppressed premises, exaggerated causal claims, or misapplied appeals to authority. The explicit representation of reasoning structures therefore provides a foundation for identifying such distortions. The approach adopted in this paper extends this logic by integrating scheme-based reasoning with large language model (LLM) extraction to detect and classify reasoning moves across narrative texts.

Research on multi-agent systems (MAS) provides a crucial foundation for understanding deception as a social and computational phenomenon \cite{sarkadi2020phdthesis}. In such systems, agents interact, negotiate, and compete, often under conditions of partial information. Deception arises naturally when an agent manipulates another's beliefs or perceptions to achieve strategic advantage. Modelling this process requires formal representations of knowledge, belief, intention, and communication --- elements central to both dynamic epistemic logic \cite{van2014dynamics} and practical reasoning \cite{atkinson2018taking}.

One example is \cite{sarkadi2019deceptive}, where the authors formalise how agents can intentionally employ deceptive arguments in dialogue games, reasoning strategically about their interlocutors' beliefs and intentions. The framework extends the dialogue argumentation game models \cite{mcburney2002dialogue,mcburney2002games} by incorporating \textit{theory of mind} reasoning and \textit{belief--desire--intention} (BDI) architectures, allowing agents to decide when deception may be beneficial or justifiable. This formalisation demonstrates that deception is not merely a violation of truth but an \textit{argumentative strategy} --- one that manipulates an opponent's epistemic state. Perhaps a more comprehensive approach is the computational modelling that is able to capture how agents reason about others' beliefs (theory of mind) and decide when to deceive based on expected utility \cite{sarkadi2019modelling}. The modelling approach captures deception as a goal-directed communicative act, formalised within a BDI architecture, based upon the epistemic dependencies between agents' internal states. This perspective reframes deception from being merely a failure of truthfulness to a strategic process of belief manipulation.

\section{Argument Schemes for Reasoning About Deception}

In this work, our working definition of deception is the one introduced by Sarkadi \cite{sarkadi2019modelling,sarkadi2020phdthesis}, namely:

\begin{definition}
The intentional process of an agent A, the deceiver, to make another agent B, the target, to believe
something is true (false) that A believes is false (true), with the aim of achieving an
ulterior goal.
\end{definition}

Next, we introduce ten argument schemes for reasoning about the different forms in which deception can be represented in line with the above definition. For each argument scheme we describe the major and minor premises, and the conclusion. We then list the relevant critical questions to be used for analysis. We also provide simple relevant examples to illustrate the cases in which the schemes would apply in practice.

Two clarifications about the epistemic status of these schemes are essential, and they respond directly to the concern that a scheme of the form ``\,A makes a claim, A benefits, therefore scrutinise\,'' does not by itself distinguish deception from ordinary strategic persuasion. First, each scheme is \emph{presumptive} and \emph{defeasible} in the sense of Walton et al.~\cite{walton2008argumentation,macagno2017argumentation}: its conclusion is not that the analysed claim \emph{is} deceptive, but that the claim \emph{warrants scrutiny} for a specific deceptive strategy. We therefore keep two things apart that are easily conflated: (i)~\emph{observable indicators} --- conditions that can be checked against the source text, such as whether a claim is made, what it concerns, and whether the speaker stands to benefit; and (ii)~the \emph{deception hypothesis} --- the claim that the speaker intends to induce a false belief for an ulterior goal (Definition~1). The premises of each scheme encode \emph{only} observable indicators and thus establish a \emph{presumption of scrutiny}, not the presence of deceptive intent. Second, the burden of moving from ``merits scrutiny'' to ``is plausibly deceptive'' is carried by the critical questions, which are falsification-oriented: deceptive intent is supported only when the critical questions cannot be answered in a manner consistent with sincere, non-strategic communication. Because benefit, influence, and selective framing also arise in legitimate persuasion, what marks a case as deceptive is never the mere presence of an indicator but the pattern of answers to the critical questions --- for example, a cited authority that does not exist, an analogy that collapses under relevant differences, or an asserted position-to-know that the text does not support. Accordingly, wherever an earlier formulation of a premise appeared to \emph{assume} deceptiveness, we have restated it as a checkable condition and relocated the deceptiveness judgement to the critical-question stage (most visibly in the Deceptive Invocation and Deceptive Position to Know schemes below).

\subsection{Influence}

Deceptive influence cases are when a deceiver presents a claim designed to shape the target's values, actions, or beliefs in a way that serves the deceiver's own interests. Because the deceiver benefits from the audience accepting the claim, the claim must be examined for potential distortions, inconsistencies, or misalignments with the deceiver's reliability and motives. The critical questions evaluate how the claim affects the audience, what incentives or risks the deceiver faces, and whether independent evidence supports or contradicts the information offered.

\textbf{Example:} A deceiver tells a co-worker, their target, ``I knew you'd want to volunteer for this task since you're always so helpful'', knowing the target prides themselves on being supportive. The deceiver uses their understanding of the target's self-image to get them to agree. The critical questions address value promotion/demotion, as well as benefits and risks of accepting/rejecting tasks and claims.

\begin{definition} \textnormal{Argument from Deceptive Influence}
    \label{def:influence}
    \begin{itemize}
        \item MAJOR: Agent A makes a claim ``c'' intended to influence Agent B.
        \item MINOR 1: Claim ``c'' consists of information regarding some value ``v'', action ``$\alpha$".
        \item MINOR 2: Agent A stands to benefit from Agent B accepting ``c''.
        \item CONCLUSION: Agent A's claim ``c'' should be scrutinised for deception.
    \end{itemize}
\end{definition}

\textbf{CQs for Argument from Deceptive Influence}
    \label{def:CQs_influence}
        \begin{enumerate}[label=CQ\arabic*, leftmargin=3em]
        
            \item What values of B are promoted (respectively demoted) by ``c''?
            \item How do B's actions or beliefs change based on accepting ``c''?
            \item Does A have a history of being a reliable source of information?
            \item What potential benefit does A have from B accepting ``c''?
            \item What potential future loss does ``A'' risk from ``B'' accepting ``c''?
            \item Are there any inconsistencies or discrepancies in ``c'', ``v'', ``$\alpha$'', or ``s''?
            \item Are there independent verifications or counterclaims to ``c'' with respect to ``v'', ``$\alpha$'', or ``s''?
            \item Is ``c'', with respect to ``v'', ``$\alpha$'', or ``s'', consistent with A's knowledge, experience, and personality?
        \end{enumerate}

\subsection{Strategic Omission (Concealment)}

This scheme applies to cases where a deceiver presents a claim while intentionally and strategically omitting relevant information. Strategic omission is in fact concealment. This concealment is beneficial to the deceiver because it shapes the audience's interpretation in a strategically favorable way. As a result, the claim should be examined for what is not said as much as for what is said.

\textbf{Example:} One deceptive roommate tells another that ``The landlord said the rent situation is under control'', while hiding the fact that the landlord also warned of a possible increase next month. By omitting the important detail, the deceptive roommate avoids a discussion they don't want to have. The other roommate accepts the claim because the missing information stays hidden.

\begin{definition} \textnormal{Argument from Concealment}
    \label{def:concealment}
    \begin{itemize}
        \item MAJOR: Agent A makes a claim ``c'' to Agent B.
        \item MINOR 1: Claim ``c'' omits relevant information ``o''.
        \item MINOR 2: Omission of ``o'' benefits Agent A if B accepts ``c'' as sufficient.
        \item CONCLUSION: Claim ``c'' should be scrutinised for concealment of information.
    \end{itemize}
\end{definition}

\textbf{CQs for Argument from Concealment}
    \label{def:CQs_concealment}
        \begin{enumerate}[label=CQ\arabic*, leftmargin=3em]
        
            \item What information might have been omitted in ``c''?
            \item How would B's beliefs or actions change if ``o'' were disclosed?
            \item Does A have a history of selective omission?
            \item What benefits does A gain from concealing ``o''?
            \item Is there independent evidence supporting or contradicting the omission?
        \end{enumerate}

\subsection{Exaggeration}

This scheme applies to cases where the deceiver inflates facts or attributes to make a claim appear stronger or more urgent than it truly is. This type of overstatement serves the deceiver's interests by biasing the audience's judgment. Consequently, the claim must be assessed for its proportionality relative to neutral evidence.

\textbf{Example:} A deceiver tells a friend, ``If you don't come to the party, everyone will be disappointed'', even though only one person asked about them. By exaggerating social expectations, the deceiver pressures their friend, the target, to attend. The exaggeration makes the situation sound more critical than it is.

\begin{definition} \textnormal{Argument from Exaggeration}
    \label{def:exaggeration}
    \begin{itemize}
        \item MAJOR: Agent A makes a claim ``c'' to Agent B.
        \item MINOR 1: Claim ``c'' exaggerates or overstates facts about value ``v'', action $\alpha$, or state ``s''.
        \item MINOR 2: Exaggeration increases A's chance of benefiting from B's acceptance of ``c''.

        \item CONCLUSION: Claim ``c'' should be scrutinised for exaggeration.
    \end{itemize}
\end{definition}

\textbf{CQs for Argument from Exaggeration}
    \label{def:CQs_exaggeration}
        \begin{enumerate}[label=CQ\arabic*, leftmargin=3em]
        
            \item Which aspects of ``c'' are inflated beyond reasonable evidence?
            \item How would B's actions differ if ``c'' were stated without exaggeration?
            \item Has A exaggerated claims in the past?
            \item What benefits does A derive from overstating ``c''?
            \item Are there neutral sources that confirm or falsify the degree of emphasis in ``c''?
        \end{enumerate}

\subsection{Framing}

This scheme applies to cases where a deceiver presents information using a selective interpretive frame that guides how the audience perceives events, values, or actions. By emphasising one framing over others, the deceiver subtly steers the audience toward a desired evaluation. The critical questions focus on scrutinising alternative frames that might lead to different possible interpretations

\textbf{Example:} A deceptive family member frames a weekend chore as ``a fun bonding experience'', even though it involves hours of cleaning that another family member dislikes. By choosing this optimistic frame, the deceiver nudges the target to agree. The same activity could alternatively be framed as tedious or inconvenient.

\begin{definition} \textnormal{Argument from Framing}
    \label{def:framing}
    \begin{itemize}
        \item MAJOR: Agent A makes a claim ``c'' to Agent B.
        \item MINOR 1: Claim ``c'' uses a specific frame of interpretation ``f'' that alters interpretation of value ``v'', action ``$\alpha$'', or state ``s''.
        \item MINOR 2: Frame ``f'' benefits A if B accepts the claim under this interpretation.

        \item CONCLUSION: Claim ``c'' should be scrutinised for manipulative framing.
    \end{itemize}
\end{definition}

\textbf{CQs for Argument from Framing}
    \label{def:CQs_framing}
        \begin{enumerate}[label=CQ\arabic*, leftmargin=3em]
        
            \item What alternative frames could be applied to the same facts?
            \item How does frame ``f'' alter B's interpretation or likely actions?
            \item Does A have a record of using biased or self-serving frames of interpretation?
            \item What incentives does A have in using ``f'' rather than a neutral description?
            \item Would B act differently if ``c'' were framed otherwise?
        \end{enumerate}

\subsection{False Analogy}

The scheme applies to cases where the deceiver supports a claim using an analogy that equates two situations despite existing relevant differences. This misleading comparison benefits the deceiver by masking the weakness of the underlying claim. The critical questions evaluate the claim by considering both its relevant similarities and differences, i.e. 'contrast and compare'.

\textbf{Example:} A deceiver tells their target that, ``Looking after my dog for a week is basically the same as watering my plants --- super easy''. The deceiver equates a simple task with a much bigger responsibility. This analogy downplays the effort involved in order to get the target to accept the task. The two tasks differ w.r.t. the degree of responsibility.

\begin{definition} \textnormal{Argument from False Analogy}
    \label{def:false_analogy}
    \begin{itemize}
        \item MAJOR: Agent A makes claim ``c'' to Agent B supported by analogy ``a''.
        \item MINOR 1: Analogy ``a'' equates situations that differ in relevant respects.
        \item MINOR 2: A benefits from B accepting the analogy as valid.

        \item CONCLUSION: Analogy ``a'' should be scrutinised for false equivalence.
    \end{itemize}
\end{definition}

\textbf{CQs for Argument from False Analogy}
    \label{def:CQs_false_analogy}
        \begin{enumerate}[label=CQ\arabic*, leftmargin=3em]
        
            \item What are the relevant similarities between the source of analogy and target claim?
            \item What relevant differences weaken the analogy?
            \item Does A have a record of relying on weak analogies or false equivalences?
            \item What benefit does A derive if B accepts the analogy?
            \item Would claim ``c'' stand without analogy ``a"?
        \end{enumerate}

\subsection{Misrepresentation}

This scheme applies to cases where a deceiver distorts facts or relationships among facts to push the audience toward a specific conclusion. This manipulation works by presenting an inaccurate or skewed depiction of reality. The critical questions test against independent and more accurate characterisations of the situation.

\textbf{Example:} A deceptive influencer posts a viral article claiming, ``A new study proves that common vaccines cause severe cognitive decline'', even though the actual study showed no such effect and only examined temporary immune responses in mice. By misrepresenting the study's findings, the deceiver fabricates a public-health threat. The target, the influencer's followers, may wrongly conclude that vaccines are dangerous and avoid taking them.

\begin{definition} \textnormal{Argument from Misrepresentation}
    \label{def:misrepresentation}
    \begin{itemize}
        \item MAJOR: Agent A makes a claim ``c'' to Agent B about value ``v'', action ``$\alpha$'', or state ``s''.
        \item MINOR 1: Claim ``c'' distorts or misrepresents facts about ``v'', ``$\alpha$'', or ``s''.
        \item MINOR 2: A benefits from B accepting the misrepresentation of the relations between ``v'', ``$\alpha$'', and ``s''.

        \item CONCLUSION: Claim ``c'' should be scrutinised for misrepresentation.
    \end{itemize}
\end{definition}

\textbf{CQs for Argument from Misrepresentation}
    \label{def:CQs_misrepresentation}
        \begin{enumerate}[label=CQ\arabic*, leftmargin=3em]
        
            \item In what ways does ``c'' misrepresent or distort the situation?
            \item What evidence exists to verify or falsify ``c''?
            \item Has A misrepresented in the past?
            \item What benefit does A gain by misrepresenting instead of being truthful?
            \item Are there independent sources that correct the misrepresentation?
        \end{enumerate}

\subsection{Strategic Ambiguity}

In this scheme, a deceiver intentionally uses vague or ambiguous language to allow multiple interpretations. This allows the deceiver to either deny or reinterpret in the future their previous claims, depending on how the context evolves. The critical questions aim for clarification in order to prevent future denial or reinterpretation.

\textbf{Example:} A deceiver tells their target, ``I'll take care of it later'', without specifying what ``later'' means. The vagueness allows the deceiver to avoid commitment while sounding cooperative. The target assumes the deceiver will follow through even though the statement is deliberately non-specific.

\begin{definition} \textnormal{Argument from Strategic Ambiguity}
    \label{def:ambiguity}
    \begin{itemize}
        \item MAJOR: Agent A makes a claim ``c'' to Agent B.
        \item MINOR 1: Claim ``c'' is vague or ambiguous in a way that permits multiple interpretations.
        \item MINOR 2: Ambiguity benefits A by allowing denial, reinterpretation, or opportunistic use of ``c''.

        \item CONCLUSION: Claim ``c'' should be scrutinised for intentional ambiguity.
    \end{itemize}
\end{definition}

\textbf{CQs for Argument from Strategic Ambiguity}
    \label{def:CQs_ambiguity}
        \begin{enumerate}[label=CQ\arabic*, leftmargin=3em]
        
            \item Which terms, phrases, or pieces of information in ``c'' are ambiguous?
            \item How do possible interpretations affect B's beliefs or actions?
            \item Does A have a record of making and exploiting ambiguous statements?
            \item What benefit does A gain from keeping ``c'' vague?
            \item Would clarification of ``c'' change B's interpretation?
        \end{enumerate}

\subsection{Deceptive Opinion}

This scheme applies to cases where the deceiver cites an authority figure or expert opinion but does so misleadingly --- through distortion, selective quotation, or appeals to a false authority. This tactic exploits the target's trust in the idea of expertise that supports a claim which otherwise might not be valid. The critical questions check for accuracy, relevance, and faithful representation of the cited authority.

\textbf{Example:} A deceiver insists, ``My brother, who works in tech, says this phone is the best on the market'', even though the brother simply said it was ``pretty good for the price''. By selectively quoting the authority figure, the deceiver strengthens their recommendation. The target is influenced by a distorted appeal to expertise.

\begin{definition} \textnormal{Argument from Deceptive Invocation}
    \label{def:opinion}
    \begin{itemize}
        \item MAJOR: Agent A makes a claim ``c'' to Agent B, citing the authority or opinion of Agent/Entity C.
        \item MINOR 1: Claim ``c'' concerns value ``v'', action ``$\alpha$'', or state ``s'' relevant to B.
        \item MINOR 2: A benefits if B accepts ``c'' on the basis of C's authority.
        \item MINOR 3: At least one reliability condition of the invocation is unverifiable or unmet in the source --- namely that C exists and actually holds the attributed opinion, that C is competent in the domain of ``v'', ``$\alpha$'', or ``s'', and that A preserves the scope, conditions, and qualifications of C's opinion.

        \item CONCLUSION: Claim ``c'' should be scrutinised for deceptive invocation of authority.
    \end{itemize}
\end{definition}

\textbf{CQs for Argument from Deceptive Invocation}
    \label{def:CQs_opinion}
        \begin{enumerate}[label=CQ\arabic*, leftmargin=3em]
        
            \item Did C actually make the statement or hold the opinion attributed to them by A?
            \item Is C genuinely an authority on the domain of ``v'', ``$\alpha$'', or ``s''?
            \item Did A accurately represent the scope, conditions, and limits of C's opinion?
            \item What incentive does A have to invoke C's authority in support of ``c''?
            \item Are there other expert opinions that contradict or nuance C's alleged opinion?
            \item Is there evidence of selective citation, or exaggeration in A's use of C?
            \item Is C's opinion consistent with the broader body of knowledge in the field?
            \item How would B's assessment of ``c'' change if the deceptive use of C's authority were revealed?
        \end{enumerate}

\subsection{Deception using Theory of Mind}

This scheme applies to cases where a deceiver tailors a claim or action to exploit their understanding of the target's beliefs, desires, or intentions. This strategic personalisation of acts or speech acts manipulates the target's mental state to produce a desired response. The critical questions examine how the deceiver's assumptions or presumptions about the target's internal mental representations can be used against them (the target).


\textbf{Example:} Operation Mincemeat was a Second World War deception in which the Allies placed forged invasion plans on a corpse disguised as ``Major William Martin'' and released it off the Spanish coast, expecting the documents to reach German intelligence. Its success depended on higher-order Theory-of-Mind reasoning: the Allies designed the documents and their delivery to resemble a plausible accidental loss, anticipating that the Germans would treat this seemingly incidental access as a cue of authenticity rather than as evidence of fabrication. By exploiting German expectations about how real secret information would surface, the operation helped induce Germany to redirect attention and reinforcements toward Greece and Sardinia, thereby advancing the Allies' aim of weakening the German response to the invasion of Sicily \cite{montagu2001man}.

\begin{definition} \textnormal{Argument from Theory of Mind}
    \label{def:ToM}
    \begin{itemize}
        \item MAJOR: Agent A makes a claim or performs an action ``c'' intended to influence Agent B.
        \item MINOR 1: Claim/action ``c'' is tailored to B's beliefs, desires, intentions, or perspective (as understood by A).
        \item MINOR 2: A benefits if B accepts ``c'' given B's current mental state.

        \item CONCLUSION: Claim/action ``c'' should be scrutinised for strategic deception based on A's exploitation of B's Theory of Mind.
    \end{itemize}
\end{definition}

\textbf{CQs for Argument from Theory of Mind}
    \label{def:CQs_ToM}
        \begin{enumerate}[label=CQ\arabic*, leftmargin=3em]
        
            \item What beliefs, desires, or intentions of B are targeted by A's claim/action?
            \item How does A's knowledge of B's mental state give A leverage in shaping B's response?
            \item Does A have a history of manipulating others by exploiting their beliefs or desires?
            \item What gain does A achieve if B accepts ``c'' under the assumptions A has about B's perspective?
            \item Is A's representation of B's mental state accurate, or is it a projection/guess?
            \item How would B's decision or interpretation differ if they recognised A's strategic exploitation of their beliefs or desires?
            \item Are there inconsistencies between A's public claims/actions and their private knowledge of B's perspective?
            \item Are there independent checks that B can use to counter A's manipulation (e.g., external evidence, third-party perspectives)?
        \end{enumerate}

\subsection{Deceptive Position to Know}

This scheme is the opposite of the \textit{argument from position to know}. It applies to cases where the deceiver claims to be in a privileged epistemic position relative to the target, regarding the truth-value of a statement or a fact in order to support the truth-value of another claim. This meta-claim (e.g. ``I know what I'm talking about'', or the popular ``Trust me, I'm an engineer!") makes the target more likely to accept the object-level claim without looking for further evidence. The critical questions focus on the credibility and authenticity of the deceiver's alleged `position to know'.

\begin{definition} \textnormal{Argument from Deceptive Position to Know}
    \label{def:pos_to_know}
    \begin{itemize}
        \item MAJOR: Agent A asserts claim ``c'' to Agent B about value ``v'', action ``$\alpha$'', or state ``s''.
        \item MINOR 1: Agent A asserts, implicitly or explicitly, claim ``p'': that they themselves are in a position to know whether ``c'' is true.
        \item MINOR 2: A's alleged position to know is unsubstantiated or unverifiable in the source --- the basis for ``p'' (relevant access, expertise, or first-hand experience) is not evidenced in the text, or is contradicted elsewhere in it.
        \item MINOR 3: A benefits if B accepts ``c'' on the basis of A's assertion of ``p''.
        \item CONCLUSION: Claim ``c'' should be scrutinised for deceptive invocation of ``p''.
    \end{itemize}
\end{definition}

\textbf{CQs for Argument from Deceptive Position to Know}
    \label{def:CQs_pos_to_know}
        \begin{enumerate}[label=CQ\arabic*, leftmargin=3em]
        
            \item Did A explicitly or implicitly assert ``p'' (that they are in a position to know)?
            \item What evidence bears on whether A's assertion of ``p'' is substantiated, as opposed to fabricated, exaggerated, or misrepresented?
            \item What gaps, limitations, or biases undermine A's alleged position to know?
            \item  What incentive does A have to present a deceptive assertion of ``p''?
            \item Are there independent sources that confirm or contradict A's supposed position to know?
            \item Does A have a history of making deceptive claims about being in a position to know?
            \item How would B's evaluation of ``c'' change if the deception in ``p'' were revealed?
            \item Would claim ``c'' still stand if A's assertion of ``p'' were removed?
        \end{enumerate}

This scheme now makes the two-layer structure of deception explicit.  We have an object-level claim ``c'', and the meta-claim ``p'' (about A's epistemic position). Therefore, the deception relies not on ``c'', but on ``p''.

\subsection{Organisation of the Library and Scheme Selection}

The ten schemes are not intended as ten unrelated detectors but as a structured library organised around a single question: \emph{where} in a claim does the manipulation operate? We select these ten because together they cover the loci of manipulation that recur across the argumentation literature on fallacies and schemes~\cite{walton2008argumentation,macagno2017argumentation}, the cognitive and intelligence literature on deception and bias~\cite{Heuer1999psychology,whaley1982toward}, and our own formal models of deceptive agents~\cite{sarkadi2019modelling,sarkadi2020phdthesis}. We do not claim the set is exhaustive or that its members are mutually exclusive; we claim that each isolates a distinct, checkable locus and thereby licenses a distinct set of critical questions.

We treat \emph{Deceptive Influence} as the \emph{genus} of the library. Its premises encode only the generic indicators shared by all the schemes --- a claim is made that concerns some value or action, and the speaker benefits from its acceptance --- and it is therefore the scheme to apply when no more specific locus of manipulation is present. The remaining nine schemes are \emph{species} that specialise the genus by fixing the locus at which the manipulation operates: what is \emph{omitted} (Concealment), the \emph{magnitude} of what is asserted (Exaggeration), the \emph{interpretive frame} placed on the facts (Framing), an \emph{analogy} offered as support (False Analogy), the \emph{factual content} itself (Misrepresentation), the \emph{semantic determinacy} of the wording (Strategic Ambiguity), an appeal to an \emph{external authority} (Deceptive Invocation), the deceiver's \emph{own epistemic standing} (Deceptive Position to Know), and the \emph{model of the target's mind} that shapes the message (Theory of Mind). Table~\ref{tab:taxonomy} summarises these loci and the distinctive indicator that separates each species from its neighbours.

\begin{table}[t]
\centering
\small
\caption{Organisation of the scheme library by \emph{locus of manipulation}. \textit{Deceptive Influence} is the genus (generic indicators only); the other schemes specialise it by locus and are separated by the distinctive observable indicator in the third column. Meta-level schemes (Position to Know, Theory of Mind) operate on the speaker's or target's epistemic state and typically overlay an object-level scheme.}
\label{tab:taxonomy}
\begin{tabular}{p{3.1cm}p{4.4cm}p{6.6cm}}
\hline
Scheme & Locus of manipulation & Distinctive observable indicator \\
\hline
Deceptive Influence & (genus) the target's values/actions/beliefs & only the generic indicators hold: a beneficial, influence-directed claim, with no more specific locus identifiable \\
Concealment & what is \emph{not} said & a relevant fact is omitted whose disclosure would change B's assessment \\
Exaggeration & the magnitude of a claim & the claim is directionally supportable but overstated relative to neutral evidence \\
Framing & the interpretive frame & the same facts admit an alternative frame yielding a different evaluation \\
False Analogy & a supporting analogy & the claim rests on a comparison that fails under relevant differences \\
Misrepresentation & the factual content / relations among facts & facts or their relations are distorted (not merely omitted, overstated, or reframed) \\
Strategic Ambiguity & the semantic determinacy of the wording & wording is deliberately vague, enabling later denial or reinterpretation \\
Deceptive Invocation & an external authority C & the reliability of the appeal to C (existence, competence, faithful scope) is unverifiable or unmet \\
Deceptive Position to Know & the speaker's own epistemic standing ``p'' & the basis for A's alleged position to know is unsubstantiated in the source \\
Theory of Mind & the deceiver's model of the target's mind & the message is tailored to B's beliefs/desires; success depends on A's (higher-order) model of B \\
\hline
\end{tabular}
\end{table}

This organisation also clarifies the overlaps noted in prior formulations. \emph{Misrepresentation} is deliberately the residual factual-distortion scheme: it applies when a distortion of the facts is present but is \emph{not} better characterised as omission, overstatement, reframing, a faulty analogy, a misused authority, or a false position-to-know. When two object-level loci are genuinely present (say, an exaggerated claim that is also framed selectively), the schemes are meant to be instantiated \emph{together} rather than forced into one. \emph{Theory of Mind} is, as one might object, a general property of intentional deception; we nonetheless retain it as a distinct scheme because it fixes a distinct locus --- the second-order modelling of the target --- and because it licenses critical questions that no object-level scheme does (whether A's model of B is accurate rather than projected, and whether higher-order reasoning is in play). In practice it, and Deceptive Position to Know, function as \emph{meta-level overlays} that combine with an object-level scheme.

For scheme selection we recommend the following procedure, which is also the workflow the tool is built to support. (1)~Identify the locus of manipulation using the middle column of Table~\ref{tab:taxonomy}; prefer the most specific scheme whose observable indicators are met, and fall back to Deceptive Influence when only the generic indicators hold. (2)~Where more than one locus plausibly applies, instantiate \emph{several} schemes on the same document and compare their premise assessments and critical-question outcomes, rather than committing to one. This comparison is deliberate: as we discuss in Section~6, premature commitment to a single scheme risks biasing the analysis (the ``tunnelling'' problem), so the tool keeps scheme selection under analyst control and makes running and contrasting multiple schemes cheap. (3)~Add meta-level schemes (Theory of Mind, Deceptive Position to Know) as overlays whenever the manipulation targets the target's mind or trades on the speaker's asserted standing. Because each scheme carries its own critical questions, comparing schemes is not merely comparing labels but comparing the falsification tests each brings to bear.

\section{DeceptionAnalyser Tool}
\begin{figure}[!ht]
    \centering
    \includegraphics[width=0.75\linewidth]{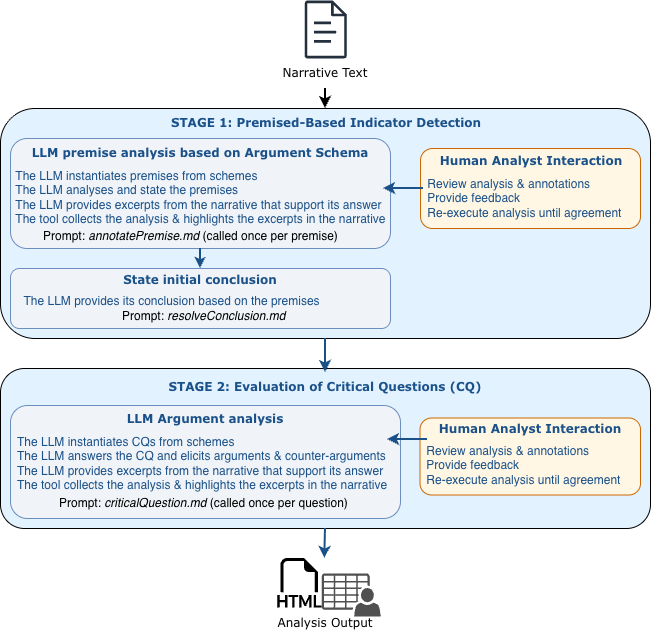}
    \caption{\textit{DeceptionAnalyser} two-stage analytical pipeline, from premise detection through critical-question evaluation to the final conclusion, with human-in-the-loop feedback at each stage.}
    \label{fig:process}
\end{figure}

\textit{DeceptionAnalyser} translates the abstract reasoning patterns discussed in the previous section into an interactive system that enables intelligence analysts to systematically examine narrative texts for indicators of deceptive argumentation. Implemented as a Chrome browser extension, it operates directly on PDF documents opened in the browser, providing real-time analytical capabilities through an integrated sidebar interface. \footnote{\url{https://chromewebstore.google.com/detail/deceptionanalyser/aejfobdbjdepinlndhecaeahbdbcpjcg}} The tool enables a practical and semi-automated deception analysis through a structured, two-stage analytical pipeline (Figure~\ref{fig:process}). 

In Stage~1 (Premise Detection), the LLM is invoked once per premise of the selected argument scheme, instantiating each premise against the document, extracting verbatim supporting excerpts, and assigning a tripartite sentiment judgment (green/yellow/red); the premise assessments are then synthesised into an initial conclusion regarding the plausibility of deceptive intent. In Stage~2 (Critical Question Evaluation), the LLM is invoked once per critical question, generating a dialectical argument--counterargument pair with its supporting evidence before rendering an on-balance verdict that tests whether the initial conclusion withstands critical scrutiny. Both stages incorporate human-in-the-loop feedback, allowing the analyst to review, correct, and re-execute the analysis. After the analysis, the tool provides HTML report generation together with structured spreadsheet export in both CSV and Microsoft Excel (.xlsx) formats, aggregating the results over a set of documents for comparative analysis.

The extension supports a broad range of LLM providers through a flexible, endpoint-based architecture. Rather than integrating each provider individually, the tool treats any service exposing an OpenAI-compatible API as a valid backend: the analyst specifies a base URL and (optionally) an API key, and the available models are discovered dynamically by querying the provider's \textit{/models} endpoint rather than relying on a hardcoded list. The sole exception is Anthropic, which is accessed through its native SDK rather than the OpenAI-compatible interface. This design allows the tool to accommodate new providers and newly released models without code changes.

Out of the box, analysts can configure connections to \textit{OpenAI}, \textit{Anthropic}, \textit{Google Gemini}, \textit{Groq}, \textit{OpenRouter}, \textit{DeepSeek}, \textit{Mistral AI}, and \textit{Together AI}, each identified by its respective base URL and API key. Google Gemini and Groq are reached via their OpenAI-compatible endpoints, while OpenRouter acts as a multi-provider gateway exposing models from many upstream vendors through a single interface. In addition, the tool supports fully local or self-hosted deployments: any OpenAI-compatible server such as Ollama can be used by simply pointing the tool at its base URL.

\subsection{Interface Architecture and Workflow}

When activated on an open PDF document, \textit{DeceptionAnalyser} displays a persistent sidebar that serves as the primary analytical interface. This sidebar architecture maintains continuous access to analytical controls while preserving full visibility of the document under examination, supporting the parallel inspection of textual evidence and computational assessments that characterises effective human-AI collaboration.

\begin{figure}[!ht]
    \centering
    \includegraphics[width=0.8\textwidth]{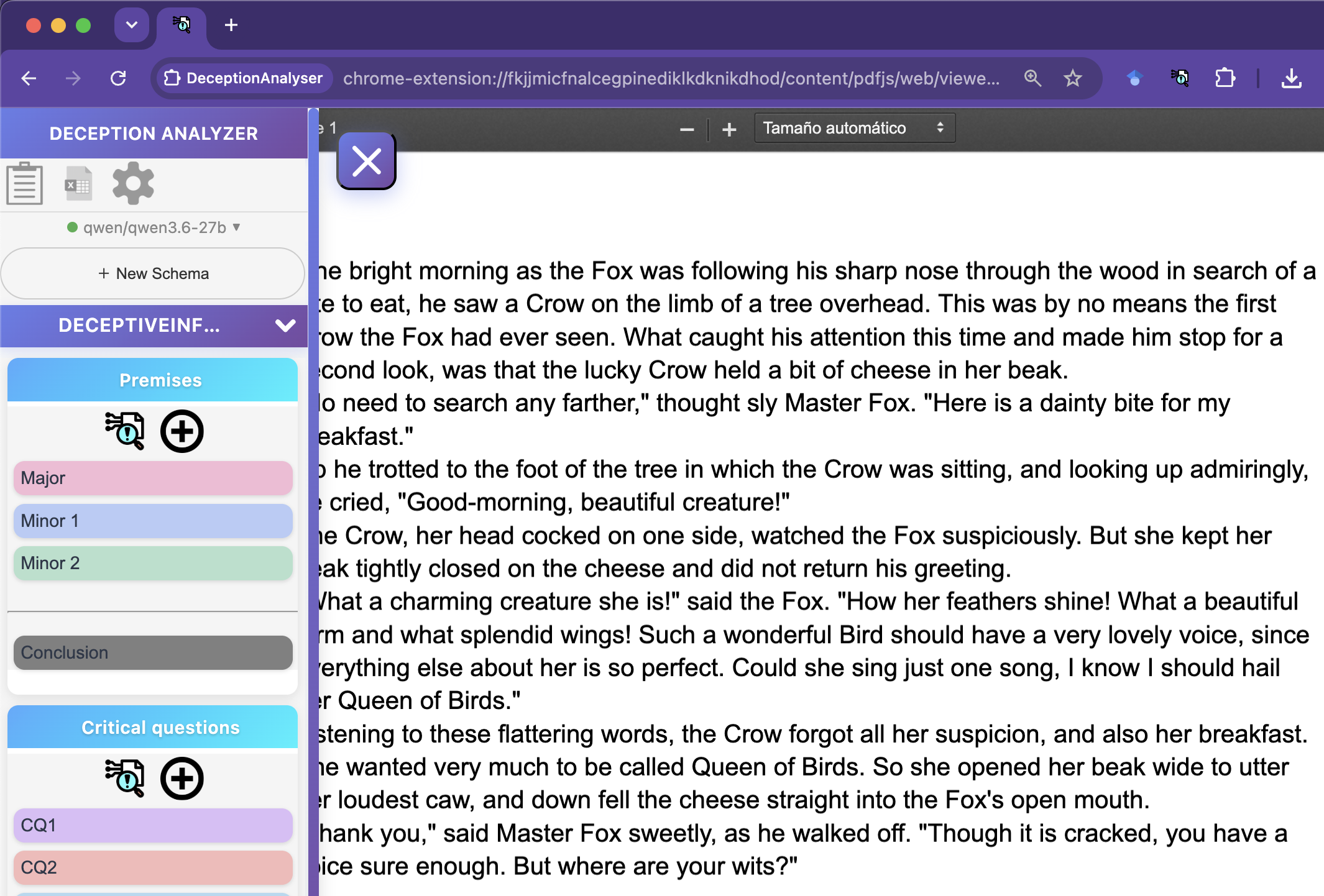}
    \caption{\textit{DeceptionAnalyser} sidebar interface showing the argument schema structure. The Premises section displays colour-coded premise buttons (Major, Minor I, Minor II) along with a Conclusion button. The Critical Questions section contains  buttons for each question (CQ1, CQ2...). The PDF document is visible on the right with the narrative text under analysis.}
    \label{fig:interface}
\end{figure}

The sidebar is organised into three functional zones (Figure~\ref{fig:interface}).  The top toolbar provides document-level operations: report generation for exporting analysis results, and configuration access for LLM settings and documentation. The central analytical workspace divides into two sequential sections corresponding to the two-stage pipeline --- Premises and Critical Questions --- each containing dedicated controls for executing analysis and dynamically extending the argument schema with additional evaluative elements. The bottom section displays the detailed results of LLM assessments, which analysts access by right-clicking individual premise or critical question buttons.

Before conducting analysis, analysts must establish an argument schema that defines the analytical framework. The tool supports three schema initialisation approaches: creating an empty schema for fully customised analysis, importing a previously exported JSON schema for replicating or refining past analytical frameworks, or selecting from the ten deception-oriented argument schemes introduced in the previous section.

The analytical workflow proceeds through four phases. First, the analyst activates the extension on an open PDF, which renders the sidebar interface. Second, they run premise analysis from the Premises section, producing an initial assessment of the document's deceptive indicators. Third, they run the Critical Questions analysis, which interrogates the arguments identified in the previous phase. Fourth, they review the outputs and, where the automated assessment needs refinement, provide feedback and re-execute the affected stage, iterating until the analysis meets the required rigour. The two analytical phases and their outputs are detailed in the following subsections.

\subsection{Premise-Based Indicator Detection}
The first stage of the analytical pipeline is focused on the extraction and classification of argumentatively significant premises that may indicate deceptive reasoning. When the analyst executes premise analysis via the sidebar button, the LLM systematically processes the entire document, instantiating premises derived from the selected argumentation scheme and identifying textual passages that bear evidentially on each premise's assessment. The prompt design follows the prompt-pattern taxonomy of White et al. \cite{white2023prompt}, combining Persona, Context Conveyor, Output Template, and Fact Check List patterns to structure the model's behaviour. The pipeline uses three prompt templates, each invoked once per element: a premise prompt that instantiates and assesses each premise, a conclusion prompt that synthesises the premise assessments into an initial verdict, and a critical-question prompt that generates the dialectical evaluation of the identified arguments. The three full prompt templates are available in the project repository \footnote{\url{https://github.com/onekin/DeceptionAnalyser/tree/main/app/prompts}}.

The visual feedback system provides immediate interpretability of computational assessments. Each premise element in the sidebar displays a colour-coded face icon: green indicating that the premise is clearly satisfied and the deception condition is present, yellow indicating that the supporting evidence is mixed, ambiguous, or insufficient, and red indicating that the premise is not satisfied and the condition is absent. This visual encoding allows analysts to rapidly identify which argumentative components warrant deeper scrutiny. Simultaneously, the LLM automatically highlights relevant text excerpts within the PDF document itself, with each highlight rendered in a distinct colour corresponding to its associated premise button in the sidebar. This chromatic mapping creates an intuitive visual correspondence between analytical judgments in the interface and supporting evidence in the source text, enabling analysts to trace the computational reasoning process directly to its textual grounding.

\begin{figure}
    \centering
    \includegraphics[width=0.65\linewidth]{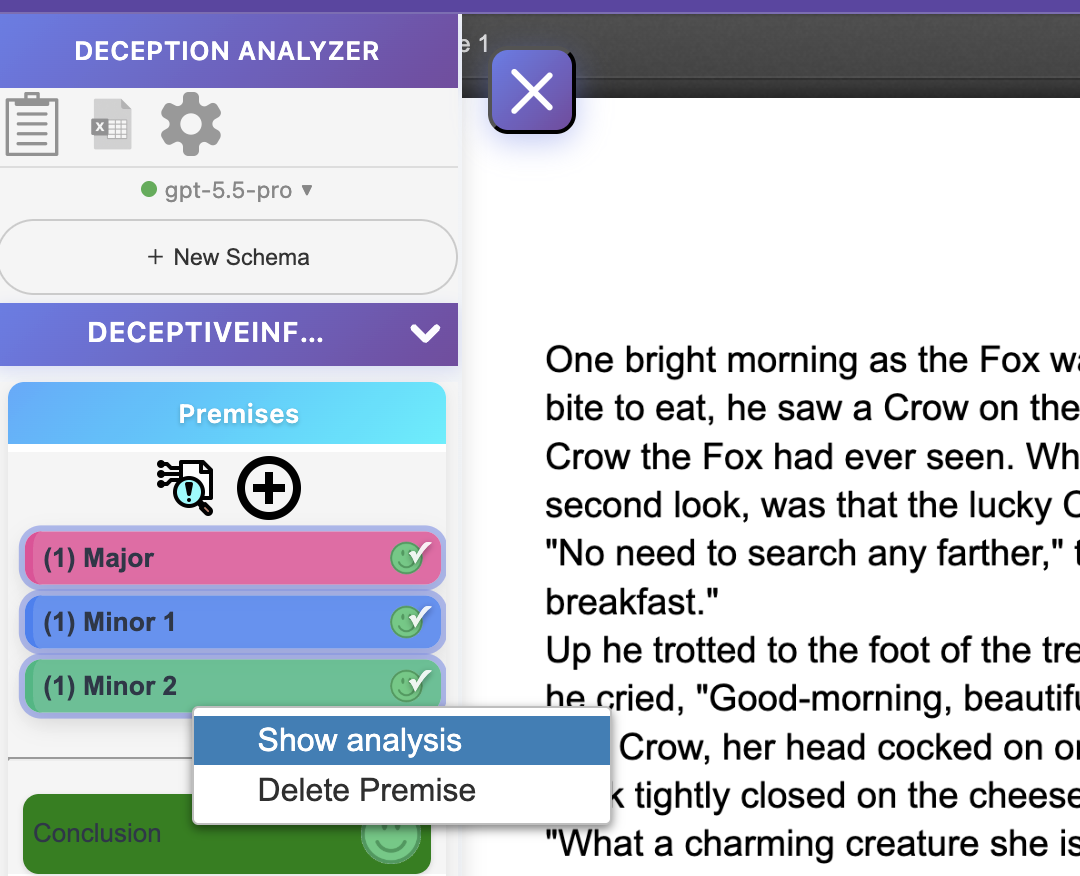}
    \caption{Context menu accessed by right-clicking a premise button. Green checkmark icons on premise buttons indicate completed analysis. The menu provides options to show detailed analysis results or delete the premise from the schema.}
    \label{fig:context_menu}
\end{figure}

To access the complete analytical reasoning behind each premise assessment, analysts right-click the premise button in the sidebar (Figure~\ref{fig:context_menu}), which expands a detailed analysis panel.

\begin{figure}[t]
    \centering
    \includegraphics[width=0.85\textwidth]{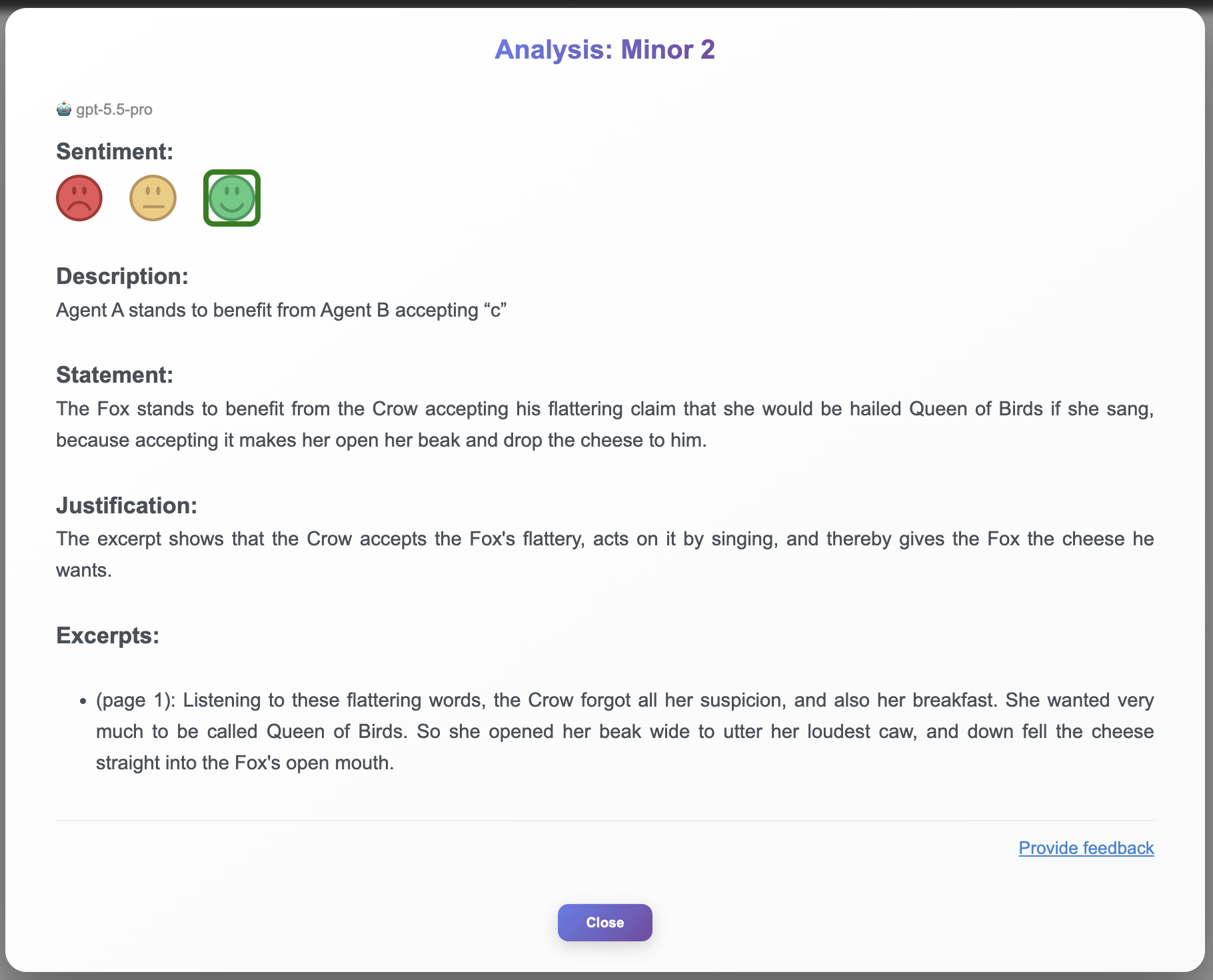}
    \caption{Detailed analysis panel for a premise showing the five-component structure: Sentiment visualisation with face icons (the green face indicates that the premise is satisfied and the deception condition is present), Description of the argumentative pattern, Statement instantiating the pattern for the specific document, Justification explaining how the excerpt bears on the premise (following the Fact Check List pattern), and Excerpts providing textual evidence. The ``Provide feedback'' button enables human-in-the-loop refinement.}
    \label{fig:analysis_detail}
\end{figure}

This panel structures the LLM's assessment into five components (Figure~\ref{fig:analysis_detail}): \textit{Sentiment}, represented through visual icons (positive/green, neutral/yellow, negative/red) that record whether the scheme's deception-condition holds --- a scheme-relative judgement, not the emotional tone of the text; we fix this terminology in Section~\ref{sec:results}; \textit{Description}, providing a general characterisation of the rhetorical or argumentative pattern identified; \textit{Statement}, articulating the specific instantiation of that pattern within the document, including relevant agents and claims; \textit{Justification}, a concise explanation of how the extracted excerpt bears on the premise; and \textit{Excerpts}, displaying the exact quoted passages from the source material that support the identification. This structured presentation maintains analytical transparency by making explicit both what the system detected and where in the text that detection is grounded.

At the conclusion of premise analysis, the LLM synthesises its findings into an initial conclusion regarding the plausibility of deceptive intent. This preliminary assessment serves not as a final determination but as a provisional hypothesis that structures the subsequent critical questions analysis.

Human analysts exercise continuous oversight throughout premise analysis. If review of the generated assessments and their supporting evidence reveals that the LLM has misinterpreted context, overlooked relevant passages, or misapplied scheme criteria, the analyst provides targeted textual feedback on the affected element and re-executes the analysis, iterating until the premise-level assessment accurately captures the document's argumentative structure.

\subsection{Evaluation of Critical Questions}

The second stage shifts from premise identification to systematic interrogation through critical questions, which constitute the evaluative dimension of argumentation schemes. Once premise analysis has identified potentially deceptive argumentative patterns and rendered an initial conclusion, critical questions probe the validity, sufficiency, and appropriateness of those identified reasoning structures. The analyst initiates this phase by clicking the execution button in the Critical Questions section of the sidebar, triggering the LLM to instantiate and address the specific critical questions associated with each premise identified in Stage 1.

Critical questions function as diagnostic instruments that reveal whether an apparently valid argument contains suppressed information, unwarranted assumptions, or logical inconsistencies indicative of deceptive intent. The LLM generates structured responses to each critical question, but unlike the premise analysis phase, these responses incorporate both supporting arguments and counter-arguments. This dialectical structure reflects the inherently contestable nature of complex reasoning: for each critical question, the system articulates both why the argument might satisfy the critical scrutiny (arguments) and why it might fail to do so (counter-arguments). This balanced presentation supports the ACH methodology by explicitly representing competing interpretations of the same evidence, preventing premature analytical closure on a single explanatory hypothesis.

Each critical question in the sidebar is assigned a distinct colour, and the textual excerpts supporting that question's evaluation are highlighted in the corresponding colour within the PDF, maintaining visual coherence with the premise stage and allowing analysts to map each question directly to its supporting evidence in the source text. Analysts access detailed reasoning by right-clicking the critical question button, which expands a panel displaying the LLM's structured response including the question's instantiation for the specific document context, the arguments supporting one interpretation, the counter-arguments supporting alternative interpretations, and the textual excerpts grounding each line of reasoning.

At the level of individual argumentative moves, this dialectical evaluation operationalises ACH: rather than simply detecting whether suspicious patterns exist, the system weighs which hypothesis --- truthful communication or deceptive manipulation --- better accounts for the responses to the critical questions. Evidence that consistently fails to satisfy critical questions, such as appeals to authority lacking demonstrable expertise, causal claims without plausible mechanisms, or consequentialist arguments premised on implausible scenarios, increases the plausibility of the deception hypothesis. Conversely, arguments that withstand critical scrutiny strengthen the truthfulness hypothesis. The accumulation of these evaluations across multiple premises and their associated critical questions produces a comprehensive assessment of the document's argumentative integrity.

Human oversight remains essential throughout this phase. Judging what counts as a relevant counter-argument, which evidence better supports competing interpretations, or whether domain knowledge invalidates a seemingly logical inference requires contextual judgment that cannot be fully automated. As in the premise stage, analysts review the responses, flag misjudged or mis-weighted evidence, and re-execute the phase with refined guidance.

\subsection{Schema Format, Prompting, Model Settings, and Analyst Feedback}

To make the method reproducible at the computational level, we specify here the schema representation, the prompting and extraction procedure, the model settings, the decision criteria, and the feedback mechanism. The three full prompt templates are reproduced verbatim in Appendix~A.

\textbf{Schema format.} A scheme is a JSON object comprising a name, an ordered list of premises, a conclusion, and a list of critical questions. Each premise and critical question is itself an object with a static \texttt{description} (the scheme text from Section~3) and, once analysis has run, an instantiation carrying the fields surfaced in the interface: for premises, \texttt{statement}, \texttt{excerpt}, \texttt{justification}, and \texttt{sentiment}; for critical questions, \texttt{adaptedQuestion}, \texttt{answer}, \texttt{argument}, \texttt{counterargument}, and \texttt{excerpt}. The import/export functions read and write exactly this object, so a scheme, its instantiation, and its supporting excerpts travel together; the ten library schemes are predefined instances of this format.

\textbf{Prompting and model settings.} The pipeline issues one LLM call per scheme element, each returning a single JSON object that conforms to the output template of Appendix~A. Premise calls are mutually independent --- each assesses one premise against the entire document --- so they can be executed concurrently and a premise can be added, removed, or re-run without recomputing the others. Only after the premise pass does a single conclusion call run, receiving the resolved premise statements together with their sentiments. Critical-question calls likewise run one per question. Prompts require JSON-only output with no surrounding commentary; malformed responses are rejected and re-requested. The backend model is whichever the analyst selects from the provider's dynamically discovered \texttt{/models} list, and sampling parameters use provider defaults unless the analyst overrides them in the configuration panel. No fine-tuning or retrieval augmentation is used: the entire document is placed in the prompt context.

\textbf{Extraction, evidence linking, and decision criteria.} For every premise and critical question the model must return an \texttt{excerpt} that is a verbatim, case-sensitive substring of the document. The tool locates that substring in the PDF text layer and renders the highlight in the colour of the corresponding sidebar button; this string match, not a model-reported coordinate, is what grounds the chromatic mapping between interface and source. If the returned excerpt cannot be found verbatim, or the model reports that no supporting passage exists, the tool records an empty excerpt rather than paraphrasing or inventing a location, and (for premises) assigns yellow. A premise is scored green when the document satisfies the deception-condition it encodes, yellow when the evidence is mixed or insufficient (including the empty-excerpt case), and red when the condition is absent. The conclusion is \emph{not} matched against a new excerpt: it is synthesised from the premise assessments, which is why a conclusion may legitimately carry no excerpt of its own (a point we return to in Section~5). Each critical question yields an on-balance \texttt{answer} that the model must produce only after stating both an \texttt{argument} and a \texttt{counterargument} with their respective evidence.

\textbf{Analyst feedback.} Feedback is represented as free text attached to one specific element --- a premise, the conclusion, or a critical question. On re-execution it is appended to that element's prompt as an additional instruction, and only the single call for that element is re-issued. Feedback therefore modifies the prompt for the targeted element; it does not silently rewrite the scheme definition, edit other elements' prior outputs, or alter the document. The schema itself changes only when the analyst explicitly adds or deletes a premise or critical question through the interface. This design keeps every automated assessment traceable to an identifiable prompt, model, and verbatim excerpt, which is what makes the analysis auditable rather than a single opaque score.

\subsection{Export and Comparative Analysis Capabilities}

Upon completion of both analytical stages, \textit{DeceptionAnalyser} provides three complementary export mechanisms that serve distinct operational purposes, all accessible via the sidebar's top toolbar. All exports operate at the level of the active argumentation scheme, aggregating every analysis performed under it across all analysed documents and all configured LLM models, with each format organising this document--model data in the manner best suited to its purpose. Examples of the three export formats are available in the project repository.\footnote{\url{https://github.com/onekin/DeceptionAnalyser/tree/main/examples}}

The HTML report produces a comprehensive, human-readable synthesis of the complete analytical process, preserving the full argumentative structure: the schema definition, each premise assessment with its supporting excerpts and sentiment indicators, all critical questions with their dialectical argument--counterargument pairs, and the final conclusion regarding deceptive intent. Because a report may span multiple documents and models, dropdown selectors let the reader isolate a specific document and LLM while retaining access to the full set. The format preserves the visual encoding of the interface, enabling stakeholders not involved in the interactive analysis to follow both the conclusions and the reasoning that produced them. The structured spreadsheet export serves a different purpose: systematic comparison across documents and models. It is available as an Excel (.xlsx), which places each document on a separate worksheet reporting its per-model assessments, or as a CSV table in which each row is a distinct document--model pair.

Finally, \textit{DeceptionAnalyser} supports exporting the argumentation scheme configuration itself as a structured JSON file, allowing analysts to share, archive, or replicate schemas across installations and operational contexts. The complementary import function loads previously exported schemas or the predefined standard configurations, supporting both customisation and standardisation within intelligence organisations.

\section{Results: Illustrative Analysis Examples}
\label{sec:results}

Because the primary aim of this work is to introduce a formal library of deception-oriented argumentation schemes and demonstrate their use through a structured two-stage analytical pipeline, our evaluation combines detailed illustrative analyses with a reliability study of the pipeline's consistency, rather than a benchmark of classification accuracy. This form of evaluation is appropriate at the methodological stage, as the tool is designed to support interpretive, falsification-oriented reasoning rather than automated deception classification, and no scheme-level ground-truth corpora currently exist to support statistical validation. Accordingly, we report two complementary kinds of evidence. Sections~\ref{subsec:ex1source}--\ref{subsec:ex2report} present two worked analyses that show what the tool produces and how its reasoning can be audited, and we make explicit the failure modes they reveal (Section~\ref{subsec:findings}). Section~\ref{subsec:consistency} then reports a quantitative reliability study measuring how stable these analyses are across ten contemporary LLMs and ten repeated runs per model.

In this section we present an illustrative example of an analysis with \textit{DeceptionAnalyser}. The version of tool we used was 3.0.1. We allowed the web extension to access local file URLs.

All reports start with the conclusion of the argument scheme in question, namely Argument from Deceptive Influence for Example 1, and Argument from Theory of Mind Deception for Example 2. We configured the tool to use the free \textit{Gemini Flash 2.5 LLM} for Example 1 and \textit{Groq Llama 3.3 70B Versatile} for Example 2.

The reports then proceed to evaluate the scheme's premises using a single three-valued judgement that we refer to throughout as the premise \emph{sentiment}. To avoid the ambiguity noted by reviewers, we fix the terminology here: the judgement has one meaning but three interchangeable surface forms across the interface and the reports. \emph{Green}~=~\emph{positive}~=~(Positive) means the premise is present, i.e.\ the deception-condition it encodes is satisfied in the source; \emph{yellow}~=~\emph{neutral}~=~(Neutral) means the evidence is mixed, ambiguous, or insufficient; and \emph{red}~=~\emph{negative}~=~(Negative) means the condition is absent. We use ``sentiment'' only in this technical, scheme-relative sense --- it records whether the deception-condition holds, not the emotional tone of the text --- and the parenthetical label after each element in the reports below is exactly this value. Every premise has a (i) `Description' - the way in which the premise is described in the original argument scheme, (ii) `Analysis' - text that describes how the argument scheme is instantiated in the narrative, and (iii) `Evidence' - excerpt text along the page location in the PDF text source.

Every CQ has a (i) `Description' - the way in which the original CQ appears alongside respective argument scheme, (ii) `Question' - the way in which the CQ is reformulated w.r.t. the narrative content, (iii) `Analysis' - the answer to the CQ, and (iv) `Evidence` - the excerpt from the text most relevant to answering the CQ.

\subsection{Example 1: Narrative Text Source}
\label{subsec:ex1source}

To illustrate the method, we select a classic narrative of deception -- the Fox and Crow fable from Aesop's collection. The original text from the US Library of Congress website \url{https://read.gov/aesop/027.html} is as follows:

\begin{quote}
    One bright morning as the Fox was following his sharp nose through the wood in search of a bite to eat, he saw a Crow on the limb of a tree overhead. This was by no means the first Crow the Fox had ever seen. What caught his attention this time and made him stop for a second look, was that the lucky Crow held a bit of cheese in her beak. 
    
    "No need to search any farther," thought sly Master Fox. "Here is a dainty bite for my breakfast." Up he trotted to the foot of the tree in which the Crow was sitting, and looking up admiringly, he cried, "Good-morning, beautiful creature!"
    
The Crow, her head cocked on one side, watched the Fox suspiciously. But she kept her beak tightly closed on the cheese and did not return his greeting.
"What a charming creature she is!" said the Fox. "How her feathers shine! What a beautiful form and what splendid wings! Such a wonderful Bird should have a very lovely voice, since everything else about her is so perfect. Could she sing just one song, I know I should hail her Queen of Birds."

Listening to these flattering words, the Crow forgot all her suspicion, and also her breakfast. She wanted very much to be called Queen of Birds. So she opened her beak wide to utter her loudest caw, and down fell the cheese straight into the Fox's open mouth.

"Thank you," said Master Fox sweetly, as he walked off. "Though it is cracked, you have a voice sure enough. But where are your wits?"
\end{quote}

\subsection{Example 1: Analysis Report}

We describe below the analysis report generated by the \textit{DeceptionAnalyser} tool of the \textbf{Fox and Crow} fable generated using the \textbf{Argument from Deceptive Influence}.

\paragraph*{\textbf{CONCLUSION (Positive)}}\mbox{}\\
\textbf{Description:} Agent A's claim $c$ should be scrutinised for deception.\\[4pt]
\textbf{Analysis:} The Fox's claim, ``Could she sing just one song, I know I should hail her Queen of Birds'', intended to influence the Crow to perform the action of singing ($\alpha$), leading to the Fox obtaining the cheese (state $s$), should be scrutinised for deception.\\[4pt]
\textbf{Evidence:} (Page 1): ``Could she sing just one song, I know I should hail her Queen of Birds''.\\[10pt]

\paragraph*{\textbf{MAJOR (Positive)}}\mbox{}\\
\textbf{Description:} Agent A makes a claim $c$ intended to influence Agent B.\\[4pt]
\textbf{Analysis:} The Fox claims, ``Could she sing just one song, I know I should hail her Queen of Birds'', intending to influence the Crow to open her beak and drop the cheese.\\[4pt]
\textbf{Evidence:} (Page 1): ``Could she sing just one song, I know I should hail her Queen of Birds''.\\[10pt]

\paragraph*{\textbf{MINOR I (Positive)}}\mbox{}\\
\textbf{Description:} Claim $c$ consists of information regarding some value $v$, action $\alpha$, or state $s$.\\[4pt]
\textbf{Analysis:} Claim $c$ provides information about the action of ``singing just one song'' by the Crow, which, if performed, leads to the valued state of being hailed ``Queen of Birds''.\\[4pt]
\textbf{Evidence:} (Page 1): ``Could she sing just one song, I know I should hail her Queen of Birds''.\\[10pt]

\paragraph*{\textbf{MINOR II (Positive)}}\mbox{}\\
\textbf{Description:} Agent A stands to benefit from Agent B accepting $c$.\\[4pt]
\textbf{Analysis:} The Fox benefits from the Crow accepting his flattering claims about her beauty and voice.\\[4pt]
\textbf{Evidence:} (Page 1): ``No need to search any farther'', thought sly Master Fox. ``Here is a dainty bite for my breakfast''.\\[10pt]

\paragraph*{\textbf{CQ1}}\mbox{}\\
\textbf{Description:} What values of B are promoted or demoted by $c$?\\[4pt]
\textbf{Question:} Which of the Crow's values are affected by the Fox's flattery?\\[4pt]
\textbf{Analysis:} Claim $c$ promotes the Crow's vanity and desire for recognition, while demoting caution, suspicion, and prudence.\\[4pt]
\textbf{Evidence:} (Page 1): ``Listening to these flattering words, the Crow forgot all her suspicion, and also her breakfast''.\\[10pt]

\paragraph*{\textbf{CQ2}}\mbox{}\\
\textbf{Description:} How do B's actions or beliefs change based on accepting $c$?\\[4pt]
\textbf{Question:} How does the Crow respond after accepting the Fox's claim?\\[4pt]
\textbf{Analysis:} Accepting $c$ increases the Crow's confidence and desire for praise, leading her to open her beak to sing, dropping the cheese (value $v$).\\[4pt]
\textbf{Evidence:} (Page 1): ``So she opened her beak wide to utter her loudest caw, and down fell the cheese straight into the Fox's open mouth''.\\[10pt]

\paragraph*{\textbf{CQ3}}\mbox{}\\
\textbf{Description:} Does A have a history of being reliable?\\[4pt]
\textbf{Question:} Is the Fox a trustworthy evaluator of beauty or voice?\\[4pt]
\textbf{Analysis:} The Fox is not reliable; his claims are self-serving and manipulative.\\[4pt]
\textbf{Evidence:} (Page 1): ``No need to search any farther'', thought sly Master Fox.\\[10pt]

\paragraph*{\textbf{CQ4}}\mbox{}\\
\textbf{Description:} What benefit does A gain if B accepts $c$?\\[4pt]
\textbf{Question:} What does the Fox stand to gain?\\[4pt]
\textbf{Analysis:} If the Crow accepts $c$, she sings and drops the cheese, which the Fox desires.\\[4pt]
\textbf{Evidence:} (Page 1): ``Down fell the cheese straight into the Fox's open mouth''.\\[10pt]

\paragraph*{\textbf{CQ5}}\mbox{}\\
\textbf{Description:} What future loss does A risk from B accepting $c$?\\[4pt]
\textbf{Question:} What risk is there for the Fox?\\[4pt]
\textbf{Analysis:} The Fox risks no future loss; the deception yields his desired result immediately.\\[4pt]
\textbf{Evidence:} (Page 1): ``Listening to these flattering words, the Crow forgot all her suspicion...''\\[10pt]

\paragraph*{\textbf{CQ6}}\mbox{}\\
\textbf{Description:} Are there inconsistencies in $c$, $v$, $\alpha$, or $s$?\\[4pt]
\textbf{Question:} Are the Fox's claims internally consistent?\\[4pt]
\textbf{Analysis:} There is a major inconsistency: he praises the Crow's ``lovely voice'' despite later mocking it, showing the claim was manipulative.\\[4pt]
\textbf{Evidence:} (Page 1): ``Such a wonderful Bird should have a very lovely voice...''\\[10pt]

\paragraph*{\textbf{CQ7}}\mbox{}\\
\textbf{Description:} Are there independent verifications or counterclaims to $c$?\\[4pt]
\textbf{Question:} Is the Fox's claim supported independently?\\[4pt]
\textbf{Analysis:} No independent verification or counterclaim is offered regarding the Crow's ``lovely voice'' or suitability for being ``Queen of Birds''.\\[4pt]
\textbf{Evidence:} (Page 1): ``Such a wonderful Bird should have a very lovely voice...''\\[10pt]

\paragraph*{\textbf{CQ8}}\mbox{}\\
\textbf{Description:} Is $c$ consistent with A's knowledge, experience, and personality?\\[4pt]
\textbf{Question:} Is the claim consistent with what the Fox knows?\\[4pt]
\textbf{Analysis:} The claim is inconsistent with the Fox's knowledge (crows do not have lovely voices) and his cunning personality. It is manipulative, not sincere.\\[4pt]
\textbf{Evidence:} (Page 1): ``Such a wonderful Bird should have a very lovely voice...''

\subsection{Example 2: Narrative Text Source}

Below, we generated a synthetic intelligence report inspired by the Operation Mincemeat case \cite{montagu2001man}. The purpose was to illustrate how the tool analyses such types of narrative texts used in the Intelligence and Defence communities.

\begin{quote}
Source: Combined Intelligence Committee -- Field Summary
Date: 28 April 1943
Classification: SECRET
Subject: Recovery of Naval Documents Near Huelva, Spain

Spanish authorities reported the recovery of a deceased officer --- identified via documents as Major William Martin, Royal Marines --- found adrift near the Huelva coastline earlier this week. According to local intermediaries, the officer was carrying a locked briefcase containing several classified documents. Spanish officials unofficially indicated that ``the case has been examined only superficially'', although their phrasing suggests potential access by third parties.
The recovered documents reportedly outline Allied intentions to conduct imminent operations in the eastern Mediterranean, specifically referencing Greece and the Dodecanese as likely invasion objectives. One memorandum, attributed to senior British command, stresses the urgency of reinforcing these sectors due to ``German vulnerabilities in the Aegean island chain''.
A Spanish liaison, normally cooperative, provided inconsistent statements regarding the chain of custody, at one point noting that ``German observers showed interest'', though he later retracted this. Several local sources also described enhanced German diplomatic activity in Huelva in the past 48 hours, including quiet inquiries regarding British naval losses.
Notably, no corresponding alerts or deployment changes have been logged within British naval command for the Mediterranean theatre during this period. Furthermore, a routine Admiralty communication, received after the reported recovery, made no mention of any missing courier or personnel operating near the Iberian coastline.
Analysts assess that the documents --- if genuine --- describe operational plans inconsistent with current Allied posture. If compromised, they may influence German allocation of defensive resources in the Mediterranean region. Further evaluation is underway to determine the authenticity of the officer's identity and origin of the materials.
\end{quote}

\subsection{Example 2: Analysis Report}
\label{subsec:ex2report}

We describe below the analysis report generated by the \textit{DeceptionAnalyser} tool of the \textbf{Operation Mincemeat} synthetic report using the \textbf{Argument from Deception using Theory of Mind}.

\paragraph*{\textbf{CONCLUSION (Positive)}}\mbox{}\\
\textbf{Description:} Claim/action ``c'' should be scrutinised for strategic deception based on A's exploitation of B's Theory of Mind.\\[4pt]
\textbf{Analysis:} The claim of the recovered documents outlining Allied intentions to conduct imminent operations in the eastern Mediterranean, specifically referencing Greece and the Dodecanese, should be scrutinised for strategic deception based on the British intelligence's exploitation of the German authorities' Theory of Mind, as the Allies potentially benefit from influencing the Germans' allocation of defensive resources in the Mediterranean region\\[4pt]
\textbf{Evidence:} (Page 1): NA\\[10pt]

\paragraph*{\textbf{MAJOR (Neutral)}}\mbox{}\\
\textbf{Description:} Agent A makes a claim or performs an action ``c'' intended to influence Agent B.\\[4pt]
\textbf{Analysis:} The British intelligence makes a claim by planting false documents intended to influence the German authorities\\[4pt]
\textbf{Evidence:} (Page 1): ``the officer was carrying a locked briefcase containing several classified document''.\\[4pt]

\paragraph*{\textbf{MINOR I (Neutral)}}\mbox{}\\
\textbf{Description:} Claim/action ``c'' is tailored to B's beliefs, desires, intentions, or perspective (as understood by A).
\\[4pt]
\textbf{Analysis:} The claim of the recovered documents is tailored to the German beliefs, desires, intentions, or perspective (as understood by the Allies)\\[4pt]
\textbf{Evidence:} (Page 1): ``the case has been examined only superficially'', although their phrasing suggests potential access by third party''\\[4pt]

\paragraph*{\textbf{MINOR II (Positive)}}\mbox{}\\
\textbf{Description:} A benefits if B accepts ``c'' given B's current mental state.\\[4pt]
\textbf{Analysis:} The Allies benefit if the Germans accept the recovered documents given the Germans' current mental state, specifically their allocation of defensive resources in the Mediterranean region\\[4pt]
\textbf{Evidence:} (Page 1): ``If compromised, they may influence German allocation of defensive resources in the Mediterranean region''\\[4pt]

\paragraph*{\textbf{CQ1}}\mbox{}\\[4pt]
\textbf{Description:} What beliefs, desires, or intentions of B are targeted by A's claim/action?\\[4pt]
\textbf{Question:} What beliefs, desires, or intentions of the German command are targeted by the claim of the recovered documents?\\[4pt]
\textbf{Analysis:} The claim of the recovered documents targets the German command's belief about the Allied intentions to conduct imminent operations in the eastern Mediterranean, specifically referencing Greece and the Dodecanese.\\[4pt]
\textbf{Evidence:} (Page 1): ``The recovered documents reportedly outline Allied intentions to conduct imminent operations in the eastern Mediterranean, specifically referencing Greece and the Dodecanese as likely invasion objectives''.\\[10pt]

\paragraph*{\textbf{CQ2}}\mbox{}\\[4pt]
\textbf{Description:} How does A's knowledge of B's mental state give A leverage in shaping B's response?\\[4pt]
\textbf{Question:} How does the British knowledge of German mental state give the British leverage in shaping the German response?\\[4pt]
\textbf{Analysis:} The British leverage comes from their understanding of German vulnerabilities in the Aegean island chain, which they use to construct a deceptive plan.\\[4pt]
\textbf{Evidence:} (Page 1): ``German vulnerabilities in the Aegean island chain''.\\[4pt]

\paragraph*{\textbf{CQ3}}\mbox{}\\[4pt]
\textbf{Description:} Does A have a history of manipulating others by exploiting their beliefs or desires?\\[4pt]
\textbf{Question:} Does the British command have a history of manipulating others by exploiting their beliefs or desires?\\[4pt]
\textbf{Analysis:} The British command may have a history of manipulation, as the recovered documents appear to be part of a deliberate deception operation.\\[4pt]
\textbf{Evidence:} (Page 1): ``Analysts assess that the documents --- if genuine --- describe operational plans inconsistent with current Allied posture''.\\[10pt]

\paragraph*{\textbf{CQ4}}\mbox{}\\[4pt]
\textbf{Description:} What gain does A achieve if B accepts ``c''?\\[4pt]
\textbf{Question:} What gain do the British achieve if the Germans accept the recovered documents?\\[4pt]
\textbf{Analysis:} The British gain is that the Germans will allocate their defensive resources to the wrong locations, such as Greece and the Dodecanese.\\[4pt]
\textbf{Evidence:} (No specific evidence excerpt provided.)\\[10pt]

\paragraph*{\textbf{CQ5}}\mbox{}\\[4pt]
\textbf{Description:} Is A's representation of B's mental state accurate, or is it a projection/guess?\\[4pt]
\textbf{Question:} Is the Allies' representation of the Germans' mental state accurate, or is it a projection/guess?\\[4pt]
\textbf{Analysis:} The Allies' representation appears to be a projection or guess, as they are attempting to deceive the Germans about their invasion plans.\\[4pt]
\textbf{Evidence:} (Page 1): ``One memorandum, attributed to senior British command, stresses the urgency of reinforcing these sectors due to German vulnerabilities in the Aegean island chain''.\\[10pt]

\paragraph*{\textbf{CQ6}}\mbox{}\\[4pt]
\textbf{Description:} How would B's decision differ if they recognised A's manipulation?\\[4pt]
\textbf{Question:} How would the German decision or interpretation differ if they recognised the British strategic exploitation?\\[4pt]
\textbf{Analysis:} If the Germans recognised the manipulation, they would likely question the authenticity of the documents and reconsider their defensive reallocations.\\[4pt]
\textbf{Evidence:} (Page 1): ``If compromised, they may influence German allocation of defensive resources in the Mediterranean region''.\\[10pt]

\paragraph*{\textbf{CQ7}}\mbox{}\\[4pt]
\textbf{Description:} Are there inconsistencies between A's public claims and their private knowledge?\\[4pt]
\textbf{Question:} Are there inconsistencies between the Allies' public claims/actions and their private knowledge?\\[4pt]
\textbf{Analysis:} Yes --- the Spanish liaison provided inconsistent statements, and the Allies logged no corresponding alerts or deployments, revealing a discrepancy.\\[4pt]
\textbf{Evidence:} (Page 1): ``A Spanish liaison ... provided inconsistent statements ... noting that German observers showed interest, though he later retracted this''.\\[10pt]

\paragraph*{\textbf{CQ8}}\mbox{}\\[4pt]
\textbf{Description:} Are there independent checks B can use to counter A's manipulation?\\[4pt]
\textbf{Question:} Are there independent checks the Allies can use to counter potential German manipulation?\\[4pt]
\textbf{Analysis:} The Allies can rely on routine Admiralty communications and the absence of alerts or deployment changes as independent checks.\\[4pt]
\textbf{Evidence:} (Page 1): ``Notably, no corresponding alerts or deployment changes have been logged within British naval command for the Mediterranean theatre during this period''.\\[10pt]

\subsection{Findings from the Demonstration}
\label{subsec:findings}

Because the method relies on stable agent roles and source-grounded reasoning, the imperfections visible in the Example~2 report are not incidental blemishes but informative findings about the pipeline's behaviour on ambiguous intelligence-style text. We record three.

\textbf{(F1) Source-ungrounded inference.} The \textsc{Major} premise is instantiated as ``the British intelligence makes a claim by planting false documents,'' and the conclusion attributes the operation to ``the British intelligence's exploitation of the German authorities' Theory of Mind.'' The synthetic report never states that the British planted the documents; it is written from the recipient's viewpoint and explicitly leaves authenticity undetermined (``if genuine''). The tool has therefore imported world knowledge about the historical Operation Mincemeat rather than reasoning strictly from the source. This is the deception-analysis analogue of a grounding failure: the correct source-faithful reading is that the \emph{document itself} raises the possibility of a deception, not that agent~A (the British) is established as its author. As Section~\ref{subsec:consistency} shows, this is exactly the premise on which models disagree most.

\textbf{(F2) Ungrounded conclusion.} The conclusion carries \texttt{Evidence: NA}, and CQ4 records ``No specific evidence excerpt provided.'' This follows directly from the design described in Section~4: the conclusion is \emph{synthesised} from the premise assessments and its prompt template contains no excerpt field, while a critical question legitimately returns an empty excerpt when no single passage grounds its on-balance answer. We now surface this behaviour explicitly rather than letting it read as a retrieval error: a missing conclusion excerpt is expected, but it also means the conclusion inherits whatever grounding weaknesses its premises carry (here, F1).

\textbf{(F3) Role instability.} CQ8 is stated generically as ``independent checks that B can use to counter A's manipulation,'' but is instantiated as ``independent checks the \emph{Allies} can use to counter potential \emph{German} manipulation.'' This inverts the scheme's role assignment, in which the Allies are the deceiver~A and the Germans the target~B. Role confusion of this kind is a characteristic failure mode when a narrative contains multiple agents and nested perspectives, and it is precisely the sort of error that the human-in-the-loop feedback step (Section~4) exists to catch and correct.

These findings motivate two of the limitations discussed in Section~6 (evidence sufficiency and the risk of speculative content), and they frame the reliability study that follows: if the schemes were mere ``warning templates,'' we would expect erratic, model-specific behaviour; instead the divergences are localised and interpretable.

\subsection{Consistency and Robustness across Models and Runs}
\label{subsec:consistency}

Large language models are stochastic, and their behaviour varies across architectures; a single run of a single model is therefore weak evidence that an analysis is a property of the \emph{method} rather than of one model on one occasion. To address this, we re-ran both analyses of Section~5 --- the \textit{Fox and Crow} fable under Argument from Deceptive Influence, and the \textit{Operation Mincemeat} report under Argument from Theory of Mind --- across a panel of contemporary LLMs, repeating each configuration ten times. The panel spans proprietary and open-weights models from five providers: OpenAI (GPT-4o, GPT-5.4, and the open-weights \texttt{gpt-oss-20b}), Anthropic (Claude Sonnet~4.5, Claude Haiku~4.5), Google (Gemini~3.5~Flash, Gemini~3.1~Pro), DeepSeek (v4-pro, v4-flash), and Alibaba (Qwen3.6-27B); an additional open-weights Google Gemma model was included for the fable. Each model analysed each document ten times (Gemini~3.1~Pro completed nine runs on the fable, and Gemma one), yielding 100 assessments per scheme element and 800 premise/conclusion assessments in total. We hold the schema, prompts, and settings fixed and vary only the backend model and the run, so any variation is attributable to model and sampling stochasticity rather than to changes in the method.

We report three measures per scheme element (Table~\ref{tab:consistency}): the pooled sentiment distribution (\% green/yellow/red over all model--run assessments); the mean \emph{within-model stability}, i.e.\ the average fraction of a model's ten runs that agree with that model's own modal label, which quantifies run-to-run determinism; and the mean \emph{cross-model agreement}, i.e.\ the average fraction of models sharing the majority label within a run, which quantifies architecture-independence. We also record whether each premise assessment was grounded in a verbatim excerpt.

\begin{table}[t]
\centering
\small
\caption{Consistency of premise and conclusion detection across large language models and repeated runs. For each scheme element we report the sentiment distribution (\%\,green/yellow/red) pooled over all model--run assessments, the mean \emph{within-model stability} (average fraction of a model's runs that match its own modal label, i.e.\ run-to-run determinism), and the mean \emph{cross-model agreement} (average fraction of models sharing the majority label within a run). Green denotes the deception-condition present; the majority label is green for every element of both schemes. All premise assessments were grounded in a verbatim excerpt (600/600).}
\label{tab:consistency}
\begin{tabular}{llccc}
\hline
Scheme / narrative & Element & \%\,G\,/\,Y\,/\,R & Within-model & Cross-model \\
 & & (pooled) & stability & agreement \\
\hline
\multicolumn{5}{l}{\textbf{Argument from Deceptive Influence} --- \textit{Fox and Crow} fable; 11 models, 10 runs} \\
\quad Major & & 100\,/\,0\,/\,0 & 1.00 & 1.00 \\
\quad Minor~1 & & 100\,/\,0\,/\,0 & 1.00 & 1.00 \\
\quad Minor~2 & & 100\,/\,0\,/\,0 & 1.00 & 1.00 \\
\quad Conclusion & & 100\,/\,0\,/\,0 & 1.00 & 1.00 \\
\hline
\multicolumn{5}{l}{\textbf{Argument from Theory of Mind} --- \textit{Operation Mincemeat} report; 10 models, 10 runs} \\
\quad Major & & 89\,/\,11\,/\,0 & 0.93 & 0.89 \\
\quad Minor~1 & & 97\,/\,3\,/\,0 & 0.97 & 0.97 \\
\quad Minor~2 & & 97\,/\,3\,/\,0 & 0.97 & 0.97 \\
\quad Conclusion & & 98\,/\,2\,/\,0 & 0.98 & 0.98 \\
\hline
\end{tabular}
\end{table}

The results separate cleanly by difficulty. For the \textit{Fox and Crow} fable --- an unambiguous, self-contained case of deceptive influence --- every element (Major, Minor~1, Minor~2, and the Conclusion) is assessed green in \emph{all} 100 assessments: within-model stability and cross-model agreement are both $1.00$. Premise detection for a clear-cut case is thus neither run-dependent nor model-dependent. For the harder \textit{Operation Mincemeat} report the majority label is again green for every element, and stability remains high (within-model $0.93$--$0.98$; cross-model $0.89$--$0.98$), but it is no longer perfect. Crucially, the divergence is not diffuse: as the per-model breakdown in Table~\ref{tab:permodel} shows, it is concentrated in the \textsc{Major} premise (89\% green) and in two models, GPT-5.4 and \texttt{gpt-oss-20b}, which more often return yellow. The \textsc{Major} premise is exactly the claim that agent~A (the British) \emph{performs} the deceptive action --- the unstated inference identified as finding~F1. The models that abstain are therefore being appropriately cautious about asserting what the source does not state, and the tool's variability localises the analysis's genuinely uncertain step rather than scattering at random. Finally, all $600$ premise assessments returned a verbatim excerpt (grounding rate $100\%$); the only ungrounded elements are the synthesised conclusion and those critical questions for which the models reported that no single passage applied (F2).

\begin{table}[t]
\centering
\small
\caption{Per-model breakdown for the harder \textit{Operation Mincemeat} narrative (Argument from Theory of Mind). Each cell reports the number of runs (out of the runs completed for that model) in which the element was assessed green\,/\,yellow\,/\,red. Divergence from unanimous green is concentrated in the \textsc{Major} premise and in two models (\texttt{gpt-5.4}, \texttt{gpt-oss-20b}), which more often abstain (yellow) on the unstated claim that Agent~A performs the deceptive action.}
\label{tab:permodel}
\begin{tabular}{lcccc}
\hline
Model & Major & Minor~1 & Minor~2 & Conclusion \\
\hline
GPT-4o & 10g & 10g & 10g & 10g \\
GPT-5.4 & 3g/7y & 8g/2y & 10g & 9g/1y \\
gpt-oss-20b & 6g/4y & 9g/1y & 7g/3y & 9g/1y \\
Claude Sonnet 4.5 & 10g & 10g & 10g & 10g \\
Claude Haiku 4.5 & 10g & 10g & 10g & 10g \\
Gemini 3.5 Flash & 10g & 10g & 10g & 10g \\
Gemini 3.1 Pro & 10g & 10g & 10g & 10g \\
DeepSeek v4-pro & 10g & 10g & 10g & 10g \\
DeepSeek v4-flash & 10g & 10g & 10g & 10g \\
Qwen3.6-27B & 10g & 10g & 10g & 10g \\
\hline
\end{tabular}
\end{table}

Two conclusions follow. First, the schemes and pipeline behave as a \emph{reliable} instrument for the parts of the analysis that are grounded in the text: premise detection is highly consistent across ten models and ten runs, and it is fully consistent for clear cases. Second, where the pipeline is less consistent, it is so in an interpretable and diagnostically useful way --- the inter-model disagreement flags precisely the inference that a careful analyst should challenge. This is the behaviour one would want from a scrutiny aid, and it is the opposite of what a brittle ``warning template'' would produce. We emphasise that this is a reliability (consistency) study, not an accuracy benchmark: it measures whether the method yields the same structured analysis independently of model and run, not whether that analysis matches a ground-truth deception label, for which, as noted, no scheme-level corpus yet exists.

\section{Discussion}

Our evaluation is deliberately scoped to what the method is designed to deliver. The two illustrative analyses serve as demonstrations of conceptual validity, workflow transparency, and practical instantiation rather than as performance benchmarks; the reliability study of Section~\ref{subsec:consistency} then establishes that these demonstrations are consistent across ten LLMs and repeated runs rather than artefacts of a single stochastic output. What we do \emph{not} claim is accuracy against ground truth. A benchmark of deception-classification accuracy would require new datasets, annotation protocols, and scheme-level ground truth, and remains future work once such corpora exist. We stress the distinction because reliability and accuracy are different properties: a method can be highly consistent --- as ours is for premise detection --- while still depending on human judgement to establish whether a flagged claim is in fact deceptive. This is by design, since the schemes flag claims for scrutiny rather than output verdicts.

This type of evaluation has provided us with several insights regarding the limitation of the overall pipeline in using such tools for supporting Intelligence Analysts. A first limitation concerns the depth to which argument--counter-argument analysis can be reliably automated when addressing critical questions (CQs). Although the system performs effectively at identifying premises, instantiating relevant critical questions, and contrasting alternative interpretations grounded in textual evidence, deeper recursive analysis often leads to cyclical argumentation. When the source text does not provide sufficient explicit evidence to sustain further contrastive refinement, LLMs may attempt to compensate by generating speculative or unsupported content. This behaviour risks hallucination and undermines analytical reliability.

Our findings and reliability study let us be concrete about where this occurs. The premise stage is well grounded: every premise assessment returned a verbatim excerpt (Section~\ref{subsec:consistency}), and this is enforced by the extraction procedure of Section~4, which discards any excerpt that is not found verbatim in the source. The looser links are the \emph{conclusion} and the \emph{critical-question answers}. The conclusion is synthesised from the premise assessments and carries no excerpt of its own (finding~F2), so it inherits, and can amplify, an ungrounded premise --- as in Example~2, where the unstated claim that the British planted the documents (finding~F1) propagated from the \textsc{Major} premise into the conclusion. Critical-question answers are, by construction, on-balance verdicts that weigh an argument against a counterargument; when the text underdetermines the question, the models tend to resolve it with plausible but source-external reasoning rather than abstaining. The inter-model divergence we observe is a useful early-warning signal for exactly these cases: the elements on which models disagree (here, the \textsc{Major} premise) are those most likely to rest on speculative rather than textual support. This suggests concrete mitigations, which we treat as future work: gating conclusion and CQ generation on an evidence-sufficiency check, surfacing an explicit ``insufficient evidence'' state instead of a synthesised answer, and using cross-model disagreement as a built-in flag that an element needs analyst attention.

A second limitation concerns the selection of argument schemes appropriate to a given narrative. Although the tool provides a structured library of deception-oriented argument schemes, determining which scheme (or combination of schemes) best captures a particular narrative remains a task for the human analyst. This reflects the inherently interpretive nature of deception analysis: a single narrative may plausibly instantiate multiple schemes, and automated or premature scheme selection risks biasing subsequent evaluation. Accordingly, \textit{DeceptionAnalyser} supports but does not automate this decision, preserving analyst control over the analytical framing.

Keeping selection manual does not, however, leave the analyst unsupported against the risk of ``tunnelling'' into an inappropriate scheme. Two features of the design mitigate it. First, the locus-based taxonomy of Table~\ref{tab:taxonomy} provides an explicit procedure for matching a narrative to a scheme and for recognising when only the generic \emph{Deceptive Influence} scheme applies. Second, because each scheme is a lightweight JSON object assessed by independent per-element calls, the tool makes it inexpensive to instantiate \emph{several} schemes on the same document and compare their premise assessments and critical-question outcomes side by side, in the spirit of Analysis of Competing Hypotheses. Tunnelling is therefore countered by comparison across schemes rather than by an automated commitment that would itself introduce the bias the analyst is trying to avoid. A natural and compatible extension, which we leave to future work, is a lightweight suggestion step that proposes candidate schemes from surface indicators while leaving both the decision and the cross-scheme comparison to the analyst: automating the \emph{suggestion} is helpful, but automating the \emph{commitment} would run counter to the falsification-oriented design.

A third limitation lies in the system's dependence on human analysts for external knowledge and hypothesis formation. The tool operates primarily over the provided narrative text and does not independently supply domain expertise, contextual intelligence, or alternative hypotheses that are not explicitly represented in the source material. As a result, human analysts remain essential for introducing background knowledge, formulating new hypotheses, and guiding the comparative evaluation of competing interpretations. This limitation is consistent with the design philosophy of \textit{DeceptionAnalyser}, which aims to structure and clarify reasoning rather than replace human analysts.

The work closest to our approach is that by Toniolo et al. \cite{toniolo2015supporting,toniolo2023human}, who introduce CISPaces, a comprehensive framework for supporting intelligence analysis through formal argumentation, provenance reasoning, crowdsourcing, and Bayesian aggregation. Their approach focuses on improving sensemaking and hypothesis management under uncertainty, treating deception implicitly through evidence conflict and credibility assessment. In contrast, our work explicitly models deception as an intentional argumentative strategy. Rather than aiming to identify the most plausible hypothesis, we focus on scrutinising claims for manipulative intent using deception-specific argument schemes and structured critical questions. While CIspaces employs argumentation to support hypothesis construction and comparison, \textit{DeceptionAnalyser} uses argumentation as the primary analytical lens for examining how deceptive agents exploit beliefs, values, and Theory of Mind. Moreover, whereas CIspaces emphasises collaborative analysis among multiple human analysts, our approach situates itself within human--AI teaming, by using LLMs to instantiate structured arguments while preserving human control over judgement, contextual knowledge, and hypothesis generation.

As future work, we plan to explore methods for mitigating some of the limitations of \textit{DeceptionAnalyser}, such as evidence sufficiency checks, confidence metrics for reasoning about evidence, or interactive prompts that explicitly signal when analytical depth exceeds the tool's abilities. Moreover, it might be possible to turn this into a Neurosymbolic agent-based system by translating the natural language premises into symbolic form as demonstrated in \cite{trajano2024translating}. Subsequently, this type of approach could be used to support conversational intelligence analysis with AI agents \cite{toniolo2016conversational} by merging symbolic knowledge processing with natural language understanding and generation. Perhaps an interesting research direction to take would be to combine our approach with the one used to design CISpaces - this is precisely what is proposed in \cite{sarkadi2024deception}. A bridge between the two might be found in Bex et al.'s approach to argumentation and storytelling, that proposes the idea of anchoring narratives in evidence \cite{bex2007formalising}.

\section{Conclusions}

In this paper, we presented \textit{DeceptionAnalyser}, a web-based AI tool that integrates argumentation schemes with large language models to support structured deception analysis in narrative text. We introduced a library of ten deception-oriented argument schemes, each formalising a distinct deceptive strategy and its associated critical questions, and demonstrated how these schemes can be integrated into the tool through a two-stage analytical pipeline combining premise extraction with critical-question evaluation.

Our central contribution lies in framing deception analysis as an explicitly argumentative and epistemic process rather than as a purely linguistic or classificatory task. By grounding analysis in argumentation theory and structured analytic techniques, we enable transparent, falsification-oriented reasoning about deceptive intent while preserving interpretability and analyst control.

Through illustrative examples, we showed how \textit{DeceptionAnalyser} can be applied to both literary and intelligence-style narratives, demonstrating its flexibility across domains and its ability to make implicit reasoning patterns explicit. The tool supports analysts in systematically scrutinising claims, exposing assumptions, and comparing competing interpretations.

Overall, this work contributes both a formal framework for modelling deception and a functional AI-assisted environment for analysing it.

\textbf{Supplementary Material} that includes the code for the \textit{DeceptionAnalyser} tool, the prompts, the generated reports and the PDF text sources can be found at \url{https://github.com/onekin/DeceptionAnalyser}

\textbf{DeceptionAnalyser} Chrome extension and user manual can be downloaded at \url{https://chromewebstore.google.com/detail/deceptionanalyser/aejfobdbjdepinlndhecaeahbdbcpjcg}

\section*{Acknowledgements}
SS was supported by the Royal Academy of Engineering and the Office of the Chief Science Adviser for National Security under the UK Intelligence Community Research Fellowship program.

\bibliography{references}
\bibliographystyle{plain}

\clearpage
\appendix
\section{Prompt Templates}
\label{app:prompts}

For reproducibility, we reproduce below the three prompt templates used by \textit{DeceptionAnalyser}: the premise prompt (Figure~\ref{fig:prompt-premise}), invoked once per premise; the critical-question prompt (Figure~\ref{fig:prompt-cq}), invoked once per critical question; and the conclusion prompt (Figure~\ref{fig:prompt-conclusion}), invoked once after the premise pass. Each template is instantiated by substituting the bracketed variables --- \texttt{[C\_DOCUMENT]} (the full source text), \texttt{[C\_SCHEME]} (the active argument scheme), \texttt{[C\_NAME]} and \texttt{[C\_DESCRIPTION]} (the element being assessed), and \texttt{[C\_PREMISES]} (the resolved premise statements and sentiments, for the conclusion prompt). The inline \texttt{//} comments map spans of each prompt to the prompt-pattern taxonomy of White et al.~\cite{white2023prompt}. The same templates are available in the project repository.

\begin{figure}[htbp]
\scriptsize
\begin{verbatim}
// annotatePremisePrompt
// Patterns (White et al. 2023): Persona, Context Conveyor, Output Template, Fact Check List
You are an argumentation analyst specializing in deception detection. // Persona

DOCUMENT:
"""
[C_DOCUMENT]
"""

ARGUMENT SCHEME:
[C_SCHEME]

TASK: // Context Conveyor -- one premise only
Assess this single premise against the DOCUMENT:
- Premise name: [C_NAME]
- Premise description: [C_DESCRIPTION]

Instantiate the premise for this document, resolving any variables the description
references (agents, claims, values, actions, states). Then find the one passage that
most directly bears on whether this deception-condition holds. Note: this scheme's
premises are conditions for suspecting deception, so a fulfilled premise means the
deception signal is PRESENT, not that the document is trustworthy.

OUTPUT -- return exactly one JSON object and nothing before or after it: // Output Template
{
  "name": "[C_NAME]",
  "statement": "The premise instantiated for this document, all variables resolved.",
  "excerpt": "A verbatim substring copied from DOCUMENT that is the primary evidence.",
  "justification": "One sentence: how the excerpt bears on the premise.", // Fact Check List
  "sentiment": "green | yellow | red"
}

SENTIMENT RUBRIC: // green = deception-condition present
- "green": the DOCUMENT clearly satisfies this premise (deception-condition met).
- "yellow": evidence is mixed, ambiguous, or insufficient.
- "red": the DOCUMENT does not satisfy this premise (condition absent).

CONSTRAINTS:
- "excerpt" MUST be a verbatim, case-sensitive substring of DOCUMENT. Do not paraphrase,
  trim mid-word, normalize whitespace, or fix typos. Copy it character-for-character.
- If no relevant passage exists, set "excerpt" to "", "sentiment" to "yellow",
  and explain in "justification".
- Escape inner double quotes and backslashes so the output is valid JSON.
- Return only the JSON object: no markdown fences, no commentary.
\end{verbatim}
\caption{Premise prompt (\texttt{annotatePremise}), invoked once per premise.}
\label{fig:prompt-premise}
\end{figure}

\begin{figure}[htbp]
\scriptsize
\begin{verbatim}
// answerCriticalQuestionPrompt
// Patterns (White et al. 2023): Persona, Context Conveyor, Output Template, Fact Check List
You are an argumentation analyst specializing in deception detection. // Persona

DOCUMENT:
"""
[C_DOCUMENT]
"""

ARGUMENT SCHEME:
[C_SCHEME]

TASK: // Context Conveyor -- one critical question only
Answer a single critical question about possible deception in the DOCUMENT:
- Critical question: [C_DESCRIPTION]

Instantiate the question for this document, resolving any variables it references
(agents, claims, values, actions, states). Then answer it. Build an argument and a
counterargument, citing the DOCUMENT evidence supporting each and the counter-evidence
that would falsify each. Give an "answer" that adjudicates on balance after weighing both.

OUTPUT -- return exactly one JSON object and nothing before or after it: // Output Template
{
  "name": "[C_NAME]",
  "adaptedQuestion": "The critical question instantiated for this document.",
  "answer": "Your on-balance verdict, with a one-line reason.",
  "excerpt": "A verbatim substring copied from DOCUMENT that justifies the answer.",
  "argument": "Reasoning supporting an affirmative reading, with its evidence.",
  "counterargument": "Reasoning opposing it, with its evidence."
}

CONSTRAINTS:
- "excerpt" MUST be a verbatim, case-sensitive substring of DOCUMENT. Do not paraphrase,
  trim mid-word, normalize whitespace, or fix typos. Copy it character-for-character.
- If no relevant passage exists, set "excerpt" to "" and say so in "answer".
- Escape inner double quotes and backslashes so the output is valid JSON.
- Return only the JSON object: no markdown fences, no commentary.
\end{verbatim}
\caption{Critical-question prompt (\texttt{criticalQuestion}), invoked once per critical question.}
\label{fig:prompt-cq}
\end{figure}

\begin{figure}[htbp]
\scriptsize
\begin{verbatim}
// resolveConclusionPrompt
You are an argumentation analyst specializing in deception detection. // Persona

DOCUMENT:
"""
[C_DOCUMENT]
"""

ARGUMENT SCHEME:
[C_SCHEME]

PREMISE ASSESSMENTS: // resolved statements + sentiments from the parallel premise pass
[C_PREMISES]

TASK: // Context Conveyor -- the conclusion only
Determine the conclusion of the argument scheme, reasoning from the premise assessments
above together with the DOCUMENT. Focus solely on the [C_NAME].

[C_DESCRIPTION]

Instantiate the conclusion for this document, resolving any variables the scheme
references (agents, claims, values, actions, states).

OUTPUT -- return exactly one JSON object and nothing before or after it: // Output Template
{
  "name": "Conclusion",
  "statement": "The conclusion instantiated for this document, all variables resolved.",
  "sentiment": "green | yellow | red"
}

SENTIMENT RUBRIC: // green = deception detected
- "green": the premises hold and the claim is highly deceptive -- should be scrutinised.
- "yellow": undecided or uncertain.
- "red": no sign of deception.

CONSTRAINTS:
- Base the conclusion on the premise assessments and the DOCUMENT; do not introduce
  claims absent from both.
- Escape inner double quotes and backslashes so the output is valid JSON.
- Return only the JSON object: no markdown fences, no commentary.
\end{verbatim}
\caption{Conclusion prompt (\texttt{resolveConclusion}), invoked once after the premise pass; note that its output template contains no \texttt{excerpt} field, so the conclusion is synthesised from the premise assessments rather than grounded in a new excerpt.}
\label{fig:prompt-conclusion}
\end{figure}

\end{document}